\documentstyle[amssymb,twoside]{article}

\newcommand{\E}{{\cal E}}
\newcommand{\B}{{\cal B}}
\newcommand{\rhos}{3}

\newcommand{\Manifold}{\Sigma}

{\catcode `\@=11 \global\let\AddToReset=\@addtoreset}
\AddToReset{equation}{section}

\makeatletter                                                           

\def\section{\@startsection {section}{1}{\z@}{-1.5ex plus -.5ex         
minus -.2ex}{1ex plus .2ex}{\large\bf}}                                 

\newtheorem{Theorem}   {Theorem}   [section]                            
\newtheorem{Corollary} [Theorem]   {Corollary}                          
\newtheorem{Lemma}     [Theorem]   {Lemma}
\newtheorem{Proposition} [Theorem] {Proposition}                        

\newcommand{\eq}[1]{(\ref{#1})}
\newcommand{\asmalla}{a}

\newcommand{\umac}{\gamma}
\newcommand{\metrict}{{}^3g }
\newcommand{\anglevar}{\theta}                                     

\newcommand{\R}{{\Bbb R}} 
\newcommand{\N}{\Bbb N} 
\newcommand{\bt}{\begin{Theorem}}
\newcommand{\et}{\end{Theorem}}
\newcommand{\bl}{\begin{Lemma}}
\newcommand{\el}{\end{Lemma}}
\newcommand{\bp}{\begin{Proposition}}
\newcommand{\ep}{\end{Proposition}}
\newcommand{\be}{\begin{equation}}
\newcommand{\ee}{\end{equation}}
\newcommand{\beq}{\begin{eqnarray}}
\newcommand{\eeq}{\end{eqnarray}}
\newcommand{\cU}{{\cal U}}
\newcommand{\cK}{{\cal K}}

\newcommand{\cC}{{\cal{C}}}
\newcommand{\cD}{{\cal{D}}}
\newcommand{\scri}{{\cal J}}
\newcommand{\om}{\omega}
\newcommand{\ka}{\kappa}
\newcommand{\Om}{\Omega}
\newcommand{\cH}{{\cal{H}}}

\newcommand{\cT}{{\cal T}}

\newcommand{\bc}{\begin{Corollary}}
\newcommand{\ec}{\end{Corollary}}

\newcommand{\eps}{{\epsilon}}
\renewcommand{\part}{{\partial}}

\newcommand{\ga}{{\metrict}}
\newcommand{\la}{{\lambda}}

\newcommand{\Si}{{\Sigma}}
\newcommand{\Xphi}{\phi}
\makeatother                                                            

\title{On ``asymptotically flat" space--times with \\
 $G_{2}$--invariant Cauchy surfaces}{}{}

\author{B.K. Berger\thanks{Permanent address:  Department of Physics,
Oakland University,
Rochester MI 48309. Supported in part by  an NSF grant \# PHY 89--04035
to the ITP, UCSB and NSF grants \#
PHY--9107162 and PHY--9305599  to Oakland University.
 e--mail: berger@vela.acs.oakland.edu}
\and
{P.T.\ Chru\'sciel}\thanks{
On leave  from the Institute of Mathematics of
the Polish Academy of Sciences.
Supported in part by  a KBN
grant \# 2 1047 9101, an NSF grant \# PHY 89--04035 to the ITP, UCSB,
and the Alexander  von Humboldt Foundation. 
e--mail: piotr@mpa-garching.mpg.de.
Present address:
Max Planck Institut f\"ur Astrophysik,  Karl
Schwarzschild Str.\ 1,  D 85740 Garching bei M\"unchen,  Germany.}
\and
{V. Moncrief}\thanks{Permanent address: Department of Physics,
Yale University, P.O.\ Box 6666,
New Haven CT 06511-8167. Supported in part by  an NSF grant \# PHY
89--04035 to the ITP, UCSB, and PHY--9201196 to Yale University.
e--mail: moncrief@yalph2.physics.yale.edu}
\\ \\
       Institute for Theoretical Physics\\ University of California\\
Santa Barbara, California 93106--4030 }
\begin{document}
\date{gr-qc/9404005}
\maketitle

\begin{abstract}
  In this paper we study space-times which evolve out of Cauchy data
  $(\Sigma,\metrict,K)$ invariant under the action of a
  two-dimensional commutative Lie group. Moreover
  $(\Sigma,\metrict,K)$ are assumed to satisfy certain completeness
  and asymptotic flatness conditions in spacelike directions.  We show
  that asymptotic flatness and energy conditions exclude all
  topologies and group actions except for a cylindrically symmetric
  $\R^3$, or a periodic identification thereof along the $z$--axis. We
  prove that asymptotic flatness, energy conditions and cylindrical
  symmetry exclude the existence of compact trapped surfaces. Finally
  we show that the recent results of Christodoulou and
  Tahvildar--Zadeh concerning global existence of a class of
  wave--maps imply that strong cosmic censorship holds in the class of
  asymptotically flat cylindrically symmetric electro--vacuum
  space--times.
\end{abstract}
%
%
%
%
%
\section{Introduction}
\label{Introduction}
It is widely believed that an important question in classical general
relativity is that of {\em strong cosmic censorship}, due to Penrose
\cite{Penrose}.  A mathematical formulation thereof, essentially due
to Moncrief and Eardley \cite{EM} ({\em cf.\/} also \cite{SCC,ChrCM}),
is the following:
\begin{quote}
{\em Consider the collection of initial data for, say,
vacuum or electro--vacuum space--times, with the initial data surface
$\Sigma$ being compact, or with the initial data
$(\Sigma,\metrict,K)$ 
--- asymptotically flat. For generic such data
the maximal globally hyperbolic development thereof is inextendible.}
\end{quote}
The failure of the above would mean a serious lack of predictability of
Einstein's equations, an unacceptable feature of a physical theory.

Because of the difficulty of the strong cosmic censorship problem, a full
understanding of the issues which arise in this context seems to be
completely out of reach at this stage. For this reason there is some
interest in trying to understand that question under various
restrictive hypotheses, {\em e.g.}, under symmetry hypotheses. Such a
program has been undertaken by one of us (V.M.) in \cite{EM,Moncrief},
and some further results in the spatially compact case have been
obtained in \cite{ChIM,IM,SCC,Chr2}. Here we consider the question of
strong cosmic censorship in the space of initial data
$(\Sigma,\metrict,K,A,E)$ for the electro--vacuum Einstein equations
which are invariant under the action of a two--dimensional commutative
Lie group $G_2$, and which satisfy some completeness and asymptotic
flatness conditions.

Clearly it would be desirable to analyze the strong cosmic censorship
problem with the minimal amount of restrictive conditions imposed, for
example assuming the existence of only {\em one} Killing vector. This
problem seems still out of reach at this stage. The next ``smallest''
isometry group possible $G_2$ is two--dimensional.
As discussed in Section \ref{general}, non--commutative
$G_2$'s are incompatible with (our notion of) asymptotic flatness.
This leads us to the commutative groups considered in this paper. Let
us also mention that an isometry group $G_n$ with $n=\mbox{dim}\
G_n\ge 3$ and asymptotic flatness seem to be compatible only with
metrics which, locally, are isometric to the Schwarzschild metric in
the vacuum case and to the Reissner--Nordstr\"om metric in the
electro--vacuum case.

Thus, in this paper we address the question of global properties of
maximal globally hyperbolic electro--vacuum space--times with complete
Cauchy surfaces $\Sigma$ and with ``asymptotically flat'' (in a sense
to be made precise below) Cauchy data invariant under the effective,
proper action of a commutative connected two--dimensional Lie group
$G_2$. We list all the possible topologies of $\Sigma$ and actions of
$G_2$ (Section \ref{general}). In that same Section we show that the
constraint equations and some energy conditions exclude all but the
standard cylindrically symmetric model: $G_2=\R\times U(1)$ with an
action on $\Sigma\approx \R^3$ by translations in $z$ and rotations in
the planes $z=\mbox{const}$ (or a periodic identification thereof
along the $z$--axis). [For these results we do not actually need to
assume that the matter is of electromagnetic nature, all we need are
some appropriate energy conditions.] We also show that trapped
surfaces which are either compact or invariant under $G_2$ are not
allowed when asymptotic flatness conditions and energy conditions are
imposed. Now it is folklore knowledge that electro--vacuum equations
and cylindrical symmetry reduce to a wave--map equation on $2+1$
dimensional Minkowski space--time. This is, however, not true without
some further conditions and the main point of the considerations in
that Section (and in fact one of the main points of this paper) is to
prove that our global hypotheses on the Cauchy data do indeed lead to
such a reduction.  In Section \ref{cylindrical} we briefly analyze
some global properties of cylindrically symmetric asymptotically flat
models: we prove the ``cylindrically symmetric positive energy
theorem''; we note the non--existence of vacuum or electro--vacuum
``spinning'' solutions.  In Section \ref{electrovac} we discuss the
reduced electro--vacuum equations; here again, the local form of the
final result is well known, but our emphasis is to take into account
the global aspects of the problem. In Section \ref{global} we use some
recent results of Christoudoulou and Tahvildar--Zadeh
\cite{ChTZ1,ChTZ2} to prove that strong cosmic censorship holds in the
class of electro--vacuum cylindrically symmetric space--times
considered here. This is the main result of this paper. On one hand,
it should be clear that at the heart of this assertion lie the deep
and difficult theorems of Christoudoulou and Tahvildar--Zadeh. On the
other, the global aspects of the reduction of the strong cosmic
censorship question to the corresponding wave--map problem have never
been considered in the literature in our context, and we believe that
several of the results presented here are new from this point of view.

 Let us close this Introduction with some
bibliographical remarks. The class of metrics analyzed here seems to
have been first considered by Kompaneets \cite{Kompaneets}; their
``polarized" counterpart has been first studied by Beck \cite{Beck}.
Significant steps in the ``reduction program" have been done by
Papapetrou \cite{Papapetrou} and, independently, by Kundt and Trumper
\cite{KT}.  The final reduction of the equations to a ``wave--map"
problem is essentially due to Ernst \cite{Ernst,Ernst2}.  A
description of cylindrically symmetric metrics using tools completely
different to ours can be found in \cite{Woodhouse}.

\section{Some general properties of space-times with two commuting Killing
vectors
  tangent to a Cauchy surface.}
\label{general}

Let $(\Sigma,\metrict,K)$ be Cauchy data for Einstein equations
\cite{ChY} (perhaps with matter satisfying some well behaved
equations, in which case the appropriate data for the matter fields
should also be given), thus $\Sigma$ is a three dimensional manifold
(which we assume throughout to be smooth, connected, paracompact,
Hausdorff; $\Sigma$ will also be assumed to be orientable unless
explicitly indicated otherwise), $\metrict$ is a Riemannian metric on
$\Sigma$ and $K$ is a symmetric tensor field on $\Sigma$.
$(\metrict,K)$ are assumed to satisfy the general relativistic
constraint equations \cite{ChY}.  We will moreover assume throughout
that $(\Sigma,\ga)$ is geodesically complete, and that there exist two
linearly independent 
vector fields $X_{a},\ a=1,2$ on $\Sigma$ such that
$$
{\cal L}_{X_a}\metrict={\cal L}_{X_a}K=0 \ ,
$$ where ${\cal L}$ denotes a Lie derivative.  It is well-known that
geodesic completeness of $(\Sigma,\ga)$ implies that the orbits of
$X_{a}$ are complete ({\em cf.\ e.g.\/} \cite{O'Neil}), hence,
assuming that there are no more linearly independent Killing vectors,
there exists a two-dimensional Lie group $G$ which acts effectively
and properly on $(\Sigma,\ga)$ by isometries.

Let us start by showing that an appropriate notion\footnote{The notion
  of asymptotic flatness described here coincides with that used in
  our strong cosmic censorship theorems for cylindrically symmetric
  electro--vacuum space--times, Section \ref{global}. The main
  justification for the ``reasonableness" of this definition is that
  it is compatible with a large class of nontrivial geometries. On the
  other hand it does not allow for the Schwarzschild geometry, or for
  initial data of ``hyperboloidal type" which are asymptotic to
  ``Scri" rather than to ``$i^o$".  The reader should, however, note
  that several results in this Section are proved assuming various
  considerably weaker notions of asymptotic flatness.} of asymptotic
flatness implies\footnote{We are grateful to Bernd Schmidt for useful
  discussions concerning this point.} that the Killing vectors have to
commute. For the purpose of the discussion here we shall say that
$(\Sigma,\ga)$ is asymptotically flat if 1) $(\Sigma,\ga)$ is
geodesically complete, with both $\Sigma$ and $\Sigma/G$ -- {\em not}
compact, 2) there exists a $G$--invariant subset $\cK$ of $\Sigma$
such that $\ga$ is flat on $\Sigma\setminus \cK$, with 3) $\cK/G$ ---
compact.  Under these conditions, the Lie algebra of a group $G$ of
isometries of $\ga$ has to be a subalgebra of the Lie algebra of
isometries of a flat $\R^3$.  Now it is easily seen that such algebras
are commutative when $\mbox{dim}\,G=2$ is assumed. This shows that a
non--commutative two--dimensional $G$ is incompatible with the notion
of asymptotic flatness defined above. In the remainder of this paper
we shall assume that $G$ is abelian, so that the Killing vectors $X_a$
commute.

Let $(M,g)$ be a globally hyperbolic development of the initial data on which
$G$ acts by isometries (such a development always exists if one solves
the
vacuum
equations, {\em cf.\ e.g.\/}   \cite{Chr1} or \cite{SCC}[Section 2.1]; more
generally, this result
still holds if the matter fields satisfy some well--behaved equations of
hyperbolic type). Define
$$
\tilde M=\{p\in M: \det\lambda_{ab}\neq0\},
$$
where
$$
\lambda_{ab}\equiv g(X_{a},X_{b}).
$$ By well known properties\footnote{Alternatively, this result
  follows directly from our list of topologies and actions given
  below.} of Killing vectors and group actions, $\tilde M$ is an open
dense subset of $M$ diffeomorphic to ${}^{2}\tilde M\times G$, for
some two-dimensional manifold ${}^{2}\tilde M$. On $\tilde M$ (passing
to an appropriate subset of $\tilde M$ if necessary) we can choose
coordinates $(t,\rho,x^{a})$ so that we have
$$
X_{a}^{\mu}\partial_\mu=\frac{\part}{\part x^a},\quad x^{a}=z,\ \anglevar \ ,
$$
and we can always parametrize the metric as
\begin{eqnarray}
g_{\mu\nu}dx^\mu dx^\nu & = & h_{AB} dx^A dx^B+
\la_{ab}(dx^a+M^a_Adx^A) (dx^b+M^b_Bdx^B), \label{GP.1}\\
h_{AB}dx^Adx^B & = & e^{2(\nu-\umac )}(-dt^2+d\rho^2), \label{GP.1.0}\\
\la_{ab}dx^adx^b & = & e^{2\umac } (dz+\asmalla  d\anglevar )^2+R^2
e^{-2\umac }d\anglevar ^2, \label{GP.1.1}
\end{eqnarray}
with some functions $\nu,\umac ,\asmalla ,R, 
M^a_A,$ which do
not depend upon
$x^a$. Note that we have
\be
R^2=\mbox{det}\,\la_{ab} \label{GP.2}
\ee
so that $R$ is the area density of the orbits of the isometry group.
The
constraint equations imply the following:
\be
D_j(X^i_a{P_i}^j)=- 
T_{i\mu}n^\mu X^i_a, \label{GP.3}
\ee
where $n^\mu$ is the future pointing  normal to $\Si$, $D_i$ is the covariant
derivative of $g_{ij}=\ga_{ij}$, with
$$
P_{ij}=\ga{}^{kl} K_{kl} \ga_{ij}-K_{ij}\ .
$$ [Here we have absorbed the usual \cite{ChY} constant $8\pi G/c^4$
in the definition of $T_{\alpha\beta}$.]  The Einstein equations for
this class of metrics can be found in the Appendix \ref{equations}
\cite{MT}.  From the constraint equations
(\ref{hamiltonianconstraint})--(\ref{momentumconstraint})
one derives
%
%
\begin{eqnarray}
  \frac{\part R_\pm}{\part\rho} & = & R_\pm\nu_\pm-h_\pm,
  \label{GP.3.0}\\ h_\pm & = & R[\umac ^2_\pm+\frac{e^{4\umac }}{4R^2}
  \asmalla ^2_\pm] + R e^{2(\nu-\umac )} T_{\mu\nu}
 n^\mu (n^\mu\pm m^\mu)
\nonumber \\ & & \qquad +\frac{e^{2(\nu-\umac )}}{4R} \la^{ab}c_ac_b,
\label{GP.3.01}
\end{eqnarray}
with
\begin{eqnarray}
f_\pm\equiv \part_\pm f &\equiv& \part_\rho f\pm \part_t f,\nonumber\\
c_a &\equiv&  \epsilon_{\mu\nu\rho\sigma} X^\mu_1 X^\nu_2 \nabla^\rho
X^\sigma_a =
2R
K_{ij} m^iX^j_a
\ ,
\label{GP.3.1}
\end{eqnarray}
and $\la^{ab}$ is the matrix inverse to $\la_{ab}$.
Here $m^\mu\part_\mu=m^i\part_i$ denotes the field of unit vectors tangent to
$\Si$ and normal to the orbits of $G$, and
$\nabla_\mu$ is the covariant derivative of
the space-time metric $g_{\mu\nu}$.

The case of compact $\Si$'s has been discussed in some detail in
\cite{Gowdy,Chr2,Moncrief} ({\em cf.\/} also \cite{IM,ChIM,SCC}).  It
has been pointed out to us by H.J.\ Seifert\footnote{H.J.\ Seifert,
  private communication. We are grateful to H.J.\ Seifert for several
  discussions concerning this point.} that the following
list\footnote{The following manifolds and actions are rather obvious.
  The point of the list given below is to emphasize that no other
  possibilities occur.} exhausts all the smooth, effective,
proper\footnote{The hypothesis that the action is proper will be
  automatically satisfied if we assume that the group $G$ here is the
  connected component of the group of {\em all} the isometries of
  $(\Manifold,\metrict) $.} actions by isometries of a commutative,
connected, two dimensional Lie group $G$ on a connected, smooth,
Hausdorff, three dimensional {\em non--compact} Riemannian manifold
$(\Manifold,\metrict) $ (up to automorphism of $G$ and diffeomorphism
of $\Manifold $):

\begin{enumerate}
\item
$G=\R\times U(1)$: \label{point1}
\begin{enumerate}
\item
\label{point1a}
$\Manifold =\R^3$:

 $G\times \Manifold \ni \Big(g=(a,e^{i\psi}),p=(\rho
\,e^{i\theta},z)\Big)\longrightarrow \phi_g(p)=(\rho
\,e^{i(\psi+\theta)}, z+a)$;

Here and below, whenever convenient a point $(x,y)\in\R^2$ is
represented in a standard way as $x+iy=\rho \,e^{i\theta}$.

When used without further qualifications, the notion of {\em
  cylindrical symmetry} will refer to this model.
\item
\label{point1b}
$\Manifold =\R^2\times S^1$:

$G\times \Manifold \ni \Big(g=(a,e^{i\psi}),p=(\rho,
e^{i\theta},z)\Big)\longrightarrow \phi_g(p)=(\rho,
e^{i(\psi+\theta)}, z+a)$;

This model will be referred to as the {\em cylindrically symmetric
wormhole}.
\item
\label{point1c}
$\Manifold =\{[1,\infty)\times S^1\times\R\}/\sim$, where the equivalence
relation $\sim$
identifies $(\rho=1,
e^{i\theta},z)$ with $(\rho=1,
e^{i(\theta+\pi)},z)$. It should be pointed out, however, that
the manifold $\Manifold $ here is {\em not} orientable.

The action $\phi_g(p)$ is the same as the one
in point \ref{point1a}.
\item
\label{point1d.1}
$\Manifold =S^2\times \R$;

The action here consists  of translations of the $\R$ factor of $\Manifold $
 and of rotations of $S^2$ around some fixed axis,
when $S^2$ is identified as a subset of $\R^3$ in the standard way.
\item
$\Manifold =\{B^2\times \R\}/\sim$, where $B^2$ is the closed two--dimensional
ball of radius $1$, $B^2=\{\rho e^{i\theta}, 0\le \rho\le 1\}\subset \R^2$, and
where the equivalence relation  $\sim$ identifies the points
$\rho e^{i\theta}$, $\rho = 1$ with $\rho e^{i(\theta+\pi)}$, $ \rho=1$. This
manifold is {\em
not} orientable.

The action here is the same as in point \ref{point1a} above.
\item
\label{point1e}
$\Manifold =S^1\times \R\times S^1$:

The action here is the same as in point \ref{point1b} above, except
for a supplementary $S^1$ identification of the $\R=\{ \rho\}$ factor of
$\Manifold $ in
\ref{point1b}.
\item
\label{point1d}
$\Manifold =\{[1,2]\times S^1\times\R\}/\sim$, where the equivalence
relation $\sim$
identifies $(\rho=1,
e^{i\theta},z)$ with $(\rho=1,
e^{i(\theta+\pi)},z)$, and $(\rho=2,
e^{i\theta},z)$ with $(\rho=2,
e^{i(\theta+\pi)},z)$. Similarly to the  points \ref{point1c} and
\ref{point1d.1} above,
the manifold $\Manifold $ here is {\em not} orientable.

The action $\phi_g(p)$ is the same as the one
in point \ref{point1a}.
\end{enumerate}

\item
$G=\R^2$:\label{point2}
\begin{enumerate}
\item\label{point2a}
$\Manifold =\R^3$: this is the standard action of $\R^2$ on $\R^3$ by
translations.
\item\label{point2b}
$\Manifold = S^1\times\R^2$; the action is by  translations in the
$\R^2$ factor of $\Manifold $.
\end{enumerate}
\item
$G=U(1)\times U(1)$:
\label{point3}
Here the manifolds and actions are as in points \ref{point1a}--\ref{point1c}
above,
except for a supplementary $S^1$ identification in the $\R=\{z\}$ factor of the
manifolds listed there. 
(The manifolds and actions listed in points
\ref{point1d.1}--\ref{point1d}
drop out, as a supplementary $S^1$ identification in the $\R=\{z\}$
factor would lead to compact models.)
\end{enumerate}

It is natural to assume that the initial data manifold $\Sigma$ is
orientable, and that $(\Sigma,g)$ is complete.
We shall show now that these requirements together with some
asymptotic flatness conditions and positivity conditions on the
energy--momentum tensor exclude essentially all cases above, except
the cylindrically symmetric model (or the periodic identification
thereof along the $z$--axis).  Let us first note that the geometries
of point \ref{point3} differ from those of point \ref{point1} by
trivial identifications and, in this sense, do not require separate
considerations.  Next, note that any $G$--invariant complete metric of
the geometries of points \ref{point1d.1}--\ref{point1d} and of point
\ref{point2b} above defines naturally a complete metric on the
corresponding compact model (in which the appropriate $\R$ factors
have been compactified to $S^1$).  It is clear that no such model can
be termed as asymptotically flat in any sense: indeed, it seems that a
reasonable prerequisite for a definition of asymptotic regions is to
require that the quotient manifold $\Sigma/G$ be {\em non--compact}.
Keeping in mind the requirement of orientability of $\Sigma$, it then
suffices to discuss the geometries \ref{point1a}, \ref{point1b}
and \ref{point2a}.

We shall start with the cylindrically symmetric wormhole, case
\ref{point1b} above: By way of
example, consider
the following metric on $M=\R^3\times S^1$:
\beq
ds^2 &=& -dt^2+dz^2+d\rho^2+(1+\rho^2)d\anglevar ^2,\label{NGP.1}\\
&& t,\rho,z\in(-\infty,\infty),\ \  \anglevar \in S^1
\approx [0,2\pi]\Big|_{\mbox{\scriptsize\rm mod}\, 2 \pi}\ .\nonumber
\eeq
Clearly $X_1=\frac{\part}{\part z}$ and
$X_2=\frac{\part}{\part\anglevar }$ generate an
action of $G=\R\times S^1$ on $(M,\metrict)$ by isometries. $M$ has
two asymptotically
flat ends connected by a throat, and one can replace $(1+\rho^2)$ by some
function which will preserve the overall features of the metric (\ref{NGP.1})
with the metric being exactly flat outside a set $\cU$ the projection of which
onto the orbit set $M/G$ is compact. The area function $R\equiv\sqrt{\det
g(X_a,X_b)}=\sqrt{1+\rho^2}$ satisfies
\beq
\part_t R &=& 0, \label{NGP.2.0}\\
\part_\rho R &
\rightarrow& \left\{ \begin{array}{ll}
                           -1\ , & \rho\to-\infty, \label{NGP.2}\\
                           1 \ ,& \rho \to \infty. \label{NGP.2.1}
                           \end{array}
                     \right.
\eeq
We shall try to mimic this behavior as follows: let $G=\R\times U(1)$ act by
translations in $z$ and rotations in $\anglevar $ on
$\Si=\left\{(\rho,z,\anglevar )\in \R^2\times S^1\right\}$, and let
$(\ga,K)$ be $G$-invariant initial data for a space-time metric of the form
(\ref{GP.1})--(\ref{GP.1.1}). Suppose moreover that there exist constants
$C,\eps>0$ such that on $\Si$ it holds that
\beq
& R \geq \epsilon, \label{NGP.3.0}\\
&\rho \leq - C:\ \ (m^\mu\pm n^\mu)\part_\mu R \leq - \epsilon, \label{NGP.3}\\
&\rho \geq  C:\ \ (m^\mu\pm n^\mu)\part_\mu R \geq \epsilon. \label{NGP.4}
\eeq
A metric satisfying (\ref{NGP.3.0})--(\ref{NGP.4}) will be called
``asymptotically of wormhole type".
Comparing with (\ref{NGP.2.0})--(\ref{NGP.2.1}), 
the conditions
(\ref{NGP.3.0})--(\ref{NGP.4}) do not seem to be overly stringent.
 Lemma 4.1 of \cite{Chr2} implies
immediately the following:

\begin{Proposition}\label{PNGP.1} No $C^2$ solutions of the constraint
equations satisfying (\ref{NGP.3.0})--(\ref{NGP.4})
 exist when the energy-momentum
tensor of the matter fields satisfies the inequality
\be
T_{\mu\nu} X^\mu Y^\nu \geq0 \label{NGP.5}\quad
\mbox{
for $X^\mu$ 
timelike future pointing, $Y^\mu$ 
null future pointing.}
\ee
\end{Proposition}

Although the point \ref{point1c} has been excluded from our
considerations by the requirement of orientability, let us
nevertheless point out that this geometry can also be excluded using
the constraint equations:
This follows immediately from the fact that the Cauchy surface for the
$2+1$ dimensional  wormhole discussed above
provides the universal covering space for the manifold $\Si$
described in point \ref{point1c}.
[This
example is a $2+1$ dimensional analogue of the
$\R P^2$ identified Schwarzschild throat, discussed in \cite{FSW}.]

Consider next the case described in point \ref{point2a} of our
classification, which is that of $pp$--wave metrics. Let thus $G=\R^2$
act on
$\R^3=\{z,\rho,\anglevar \in \R\}$ by translations in $z$ and $\anglevar $, and
suppose that $(\ga,K)$ are $G$-invariant initial data
such that \begin{enumerate}
\item $\ga$ is flat outside of a set $\cK$ such that $\cK/G$
  is compact;
\item   $(\Sigma,\ga)$ is  a  complete Riemannian manifold; it then
  follows that in the   coordinates of  (\ref{GP.1})--(\ref{GP.1.1})
$\rho$ covers the whole range $ (-\infty,\infty)$;
\item $\lim _{|\rho|\to\infty}\frac{\partial R}{\partial n} = 0$, where
  $\frac{\partial R}{\partial n}$ is the derivative of $R$ in the direction
  normal to $\Sigma$.
\end{enumerate}
Such initial data will be called {\em initial data for a localized
  $pp$--wave}. We have the following result, somewhat reminiscent of a
result of Penrose \cite{PenroseRMP}:

\begin{Proposition}\label{PNGP.2} Let $(\Si,\ga,K)$ be $C^2$ initial data for
  a localized $pp$--wave. If (\ref{NGP.5}) holds, then we must have
  $T_{\mu\nu}n^\mu (n^\mu\pm m^\mu)|_\Si\equiv0$, and $(\Si,\ga,K)$
  must be initial data for Minkowski space-time.
\end{Proposition}

{\bf Proof:} It is an easy exercice to find the most general form of
$\ga$ on $\Sigma\setminus \cK$. One finds in particular that we must have
$$
\lim _{\rho\to-\infty}\frac{\partial R}{\partial \rho} \le 0\,,\qquad
\lim _{\rho\to\infty}\frac{\partial R}{\partial \rho} \ge 0\,,
$$ and that the functions $f_\pm$ appearing in eq.\ (4.1) of
\cite{Chr2} satisfy $f_\pm\in L^1(\R)$. The result follows now from
the arguments of the proof of Lemma 4.1 of \cite{Chr2}.  \hfill$\Box$

Let us note, that both in Propositions \ref{PNGP.1} and \ref{PNGP.2}
we can allow initial data of $C^1$ differentiability and with a
distributional component in $T_{\mu\nu}$ (in which case we need to assume that
(\ref{NGP.5}) holds in a distributional sense).

The asymptotic flatness conditions in Proposition \ref{PNGP.2} can be
considerably weakened; this will be discussed elsewhere.

In the remainder of this paper we shall be concerned with cylindrical symmetry:
Namely, let $G= \R\times U(1)$ act on $\Si=\R^3$ by translations in $z$ and
rotations in the planes $z=\mbox{\rm const}$. Initial data $(\ga,K)$
on $\Si=\R^3$
invariant under this action of $G$ will be called cylindrically symmetric. It
is well-known that this topology of $\Si$ and this action of $G$ are compatible
with
initial data $(\Si, \ga, K)$ such that $(\ga,K)$ are data for Minkowski
space-time outside a set $\cK$ such that $\cK/G$ is compact. We have the
following generalization of Corollary~5.2 of \cite{Chr2} ({\em cf.\/} also
\cite{Thorne,Gowdy}), which follows immediately from eqs.\
(\ref{GP.3.0})--(\ref{GP.3.01}) and from the arguments of the proof of
Corollary~5.2 of \cite{Chr2}:

\bp\label{PGP.1} Let $(\Si,\ga,K)$ be $C^2$ cylindrically symmetric
initial data and suppose that (\ref{NGP.5}) holds.  Suppose moreover
that there exists $C>0$ such that \be t=0,\quad \rho\geq C:\qquad
(m^\mu\pm n^\mu)\nabla_\mu R>0. \label{GP.6} \ee [Recall that $n^\mu$
is a field of unit normals to $\Si$, and $m^\mu$ is unit, tangent to
$\Si$ and orthogonal (outwards pointing) to the orbits of $G$.] Then
\begin{enumerate}
\item $\nabla^\mu R$ is spacelike on $\Si$,
\item There are no trapped surfaces $\cT \subset\Si$ which are either
compact, or invariant under $G$.
\end{enumerate}
\ep

Note that the vector fields $m^\mu\pm n^\mu$ are null and orthogonal
to the orbits of $G$. For cylindrically symmetric metrics we shall
always choose $m^\mu$ to be outwards directed in the obvious sense,
then the above null vector fields will also be called outwards
directed.

Let us mention that (\ref{GP.6}) can be thought of as a rather mild asymptotic
flatness condition (compare eqs.~(\ref{NGP.2.0}), (\ref{NGP.2.1})). Let us also
note, that if a globally hyperbolic development $(M,g)$ of $(\Si,\ga,K)$ can
be foliated by Cauchy surfaces on which (\ref{NGP.5}) and (\ref{GP.6})
 hold, then
$\nabla^\mu R$ will be globally spacelike, and there will be no trapped
surfaces of the kind considered above in $M$. [It should be clear from
the analysis below that if $T_{\mu\nu}(n^\mu n^\nu -
m^\mu m^\nu)\equiv0$ holds, then these last conclusions will  hold
when (\ref{GP.6})
 holds on one single Cauchy surface, as long as dynamics
preserves (\ref{GP.6}). This is indeed the case for electro--vacuum
space--times satisfying apropriate asymptotic conditions, {\em
cf.\/} Theorem \ref{T4.2} below.]

To proceed further, let us rewrite somewhat more explicitly the metric
(\ref{GP.1}) in the form
\beq
g_{\mu\nu}dx^\mu dx^\nu & = & e^{2(\nu-\umac )} (-dt^2+d\rho^2)+\la_{ab}
(dx^a+M^adt+g^a d\rho)\qquad \nonumber\\
& & \qquad \qquad\qquad \quad\qquad\qquad\times (dx^b+M^b
dt+g^bd\rho).\label{GP.7}
\eeq
Replacing $x^a$ by $x^a+\int^\rho_0g^a(t,s)ds$ we can achieve
\be
g^a\equiv0. \label{GP.7.1}
\ee
[It is not too difficult to show, using {\em e.g.\/} the methods of \cite{Chr2}
[Appendix C], that the above is a smooth coordinate transformation if the
metric $g_{\mu\nu}$ is smooth.] A {\sc Mathematica} calculation (using
{\sc MathTensor} \cite{MT}) gives the equations
({\em cf.\/} (\ref{M1rhoeq})--(\ref{M2rhoeq}))
\beq
\frac{\part M^z}{\part\rho} & = &-\asmalla  \frac{e^{2\nu}}{R^3}
(\asmalla  c_1-c_2) -
\frac{e^{2\nu-4\umac }}{R} c_1, \label{GP.9}\\
\frac{\part M^\anglevar }{\part\rho} & = & \frac{e^{2\nu}}{R^3}
(\asmalla  c_1-c_2).
\label{GP.9.1}
\eeq
Let us also assume that
\be
T_{\mu\nu} n^\mu X_a^\nu\equiv0. \label{GP.10}
\ee
Equation~(\ref{GP.3})
 implies that $\part c_a/\part\rho=0$. By (\ref{GP.3.1}) 
the $c_a$'s
vanish on the axis of symmetry, hence
 it follows that ({\em cf.\/} also
\cite{KT,Geroch})
\be
c_a\equiv0. \label{GP.11}
\ee
(\ref{GP.9})--(\ref{GP.9.1}) now give $\part M^a/\part\rho\equiv0$.
We have $M^\theta|_{\rho=0}=0$
by regularity of $g_{\mu\nu}$ on the symmetry axis, and replacing $z$
by $z+\int_0^tM^z(s)ds$ leads to 
\be M^a\equiv0. \label{GP.12} \ee A {\sc Mathematica} calculation
shows that one also has ({\em cf.\/} (\ref{Requation}))
 \beq \frac{\part^2 R}{\part t^2} -
\frac{\part^2R}{\part\rho^2} &=& \frac{e^{2\nu}}{2R} \left[e^{-4\umac
  }c_1^2 + \frac{(c_2-\asmalla c_1)^2}{R^2}\right] \nonumber\\ &&
\quad + \ e^{2(\nu-\umac )} R T_{\mu\nu}(n^\mu n^\nu-m^\mu m^\nu).
\label{GP.14} \eeq Let us suppose that (\ref{GP.11}) holds and that
\be T_{\mu\nu}n^\mu n^\nu=T_{\mu\nu}m^\mu m^\nu. \label{GP.15} \ee It
follows that the right-hand-side of (\ref{GP.14}) vanishes, hence
there exist functions $f$, $g$ such that
$$
R=f(\rho+t)+ g(\rho-t).
$$
Define
\beq
\rho' &=& f(\rho+t)+g(\rho-t)\label{GP.16.0}\\
t' &=& f(\rho+t) - g(\rho-t) \label{GP.16}
\eeq
Eqs.\ \eq{B.18}--\eq{B.20} show that we have
\beq
\det \frac{\part(\rho',t')}{\part(\rho,t)} &=& 4f'(\rho+t)g'(\rho-t)
\nonumber\\
&=& e^{2(\nu-\umac )}g^{\mu\nu} R_{,\mu}R_{,\nu}.\label{GP.17}
\eeq
If we assume now that
(\ref{NGP.5}) and (\ref{GP.6}) hold, point 1. of Proposition
\ref{PGP.1} shows 
that (\ref{GP.16.0})--(\ref{GP.16}) define a
diffeomorphism in a neighborhood of $\Si$ in $M$ (the axes of symmetry, again,
can be taken care of by the methods of Appendix~C of \cite{Chr2}). Dropping
primes, in the new coordinates we have $R\equiv\rho$, so that we can put the
metric in the Kompaneets form \cite{Kompaneets}
\beq
ds^2 &=& e^{2(\nu-\umac )} (-dt^2+d\rho^2)+\la_{ab}dx^adx^b, \label{GP.18}\\
&&\det \la_{ab}=\rho^2. \label{GP.19}
\eeq
To summarize, we have proved the following:

\begin{Theorem}\label{TPG.1}
Let $(M,g)$ be a cylindrically symmetric globally hyperbolic
space--time with 
Killing vectors $X_a$, $a=1,2$, with
$M\approx (-T,T)\times \Sigma$ for some $0<T<\infty$, and with
$\Sigma\approx\R^3$. Suppose 
that the
energy--momentum tensor of $g$ 
satisfies the following:
\beq
(g^{\mu\nu}-\la^{ab}X^\mu_aX^\nu_b)X^\rho_cT_{\nu\rho} &=& 0, \label{GP.20}\\
(g^{\mu\nu}-\la^{ab}X^\mu_aX^\nu_b)T_{\mu\nu} &=& 0, \label{GP.21}
\eeq
(recall that $\la_{ab}=g(X_aX_b)$, and that $ \la^{ab}$ is the matrix
inverse to the matrix $\la_{ab}$), and
\be T_{\mu\nu}X^\mu Y^\nu\geq0 \label{GP.22} \ee for all $X^\nu$ ---
null, $Y^\nu$ --- timelike, consistently time--oriented. Suppose
moreover that there exists $C>0$ such that \be \rho\geq C : \quad
Y^\mu\nabla_\mu R>0 \label{GP.23old} \ee where $Y^\mu$ is any {\em outwards
directed} null vector orthogonal to the orbits of $G$, and $\rho $ is
any coordinate labelling the orbits of the groups on the hypersurfaces
$\{\tau\}\times\Sigma$.  Then there exists a global coordinate system
on $M$ such that $(M,g)$ is
isometric to a subset of $\R^4$
with a metric of the form
(\ref{GP.18})--(\ref{GP.19}).
\end{Theorem}

Strictly speaking,  in Theorem \ref{TPG.1} the condition \eq{GP.20}
can be replaced by the weaker condition \eq{GP.10}. 
Condition \eq{GP.20} seems to be somewhat more elegant, as 
it does not explicitly use the foliation--dependent 
vector field $m^\mu$.

As discussed in detail below, the hypotheses
(\ref{GP.20})--(\ref{GP.22}) on the energy--momentum tensor are
satisfied by cylindrically symmetric electro--vacuum space--times
({\em
cf.\/} Theorem \ref{TPG.3} below). It
should be noted that in general $i(\Si)$ will {\em not\/} be given by the
equation $t=$ const. Let us also mention that the tensor
$$
h^{\mu\nu}=g^{\mu\nu}-\la^{ab}X^\mu_aX^\nu_b
$$
which appears in (\ref{GP.20})--(\ref{GP.21}) is a projection operator
on the space orthogonal to the orbits of the isometry
group $G$.

Choosing $(M,g)$ in Theorem \ref{TPG.1}  to be (perhaps an appropriate
subset of) the maximal globally
hyperbolic \cite{ChY} development  of  $(\Si,\ga,K)$ one obtains:
\begin{Theorem}\label{TPG.2} Let $(\Si,\ga,K)$ be cylindrically symmetric
  initial data for vacuum Einstein equations. Suppose moreover that
  there exists $C>0$ such that \be x^3\geq C : \quad Y^\mu\nabla_\mu
  R\Big|_\Sigma>0 \label{GP.23} \ee where $Y^\mu$ is any {\em outwards
  directed} null vector orthogonal to the orbits of $G$, and $x^3\ge 0$
  is any coordinate parametrizing the orbits of the group on $\Sigma$.
  Then there exists a globally hyperbolic vacuum space-time $(M,g)$
  with a metric of the form (\ref{GP.18})--(\ref{GP.19}) and an
  isometric embedding $i:\Si\to M$ such that $i(\Si)$ is a Cauchy
  surface for $(M,g)$.
\end{Theorem}

{It will be seen in Section \ref{global}
that, under the conditions of  Theorem \ref{T4.2}, the
coordinates in which the metric takes the form
(\ref{GP.20})--(\ref{GP.21}) are global on the
maximal globally hyperbolic development of the data; {\em cf.\/}
Corollary \ref{C4.1}.

\section{Asymptotically flat cylindrically symmetric geometries.}
\label{cylindrical}
Let us consider a cylindrically symmetric Riemannian metric $\ga$ on
$\Si=\R^3$,
parametrized as in (\ref{GP.1})--(\ref{GP.1.1}):
\beq
\ga_{ij}dx^idx^j &=& e^{2(\nu-\umac )} d\rho^2+\la_{ab} (dx^a+g^ad\rho)
(dx^b+g^bd\rho), \label{N3.1}\\
\la_{ab}dx^adx^b &=& e^{2\umac } (dz+ \asmalla d\anglevar )^2 + R^2
e^{-2\umac }d\anglevar ^2. \label{N3.2}
\eeq
Without loss of generality, rescaling $\rho$ if necessary, we may assume
\be
e^{2\nu}|_{\rho=0}=1. \label{N3.3}
\ee
The regularity of $\ga$ at the axis $\rho=R=0$ requires that the limits
\beq
\lim_{\rho\to0} \frac{g^\anglevar }{\rho^2}, && \lim_{\rho\to0} \frac{\asmalla
}{\rho^2}
\quad {\rm exist,\ and\ are\ finite;\/} \label{N3.4}\\
&& \lim_{\rho\to0} \frac{R}{\rho}=1. \label{N3.5}
\eeq
For the purpose of this section we shall say that the initial data
$(\metrict,K)$ are
asymptotically flat if the limits
\beq
& \lim_{\rho\to\infty}\nu,\quad  \lim_{\rho\to\infty}\umac \quad
  {\rm exist,\ and\ are\ finite;\/}
\label{N3.6}\\
& \lim_{\rho\to\infty} R_\rho\quad
 {\rm exists\ and\ is\ (strictly)\ positive;\/}
\label{N3.7}\\
& \lim_{\rho\to\infty}\rho g^\anglevar =\lim_{\rho\to\infty}g^z=
\lim_{\rho\to\infty}R_t=0.\label{N3.7.1}
\eeq
[Clearly, the notion of asymptotic flatness used here is compatible with
(and stronger than) the one used in the previous Section.]
Assuming that the energy condition \eq{NGP.5} holds,
it follows from Proposition \ref{PGP.1} that
\be
R_\pm>0, \label{N3.11}
\ee
so that, using (\ref{N3.3}), we can solve (\ref{GP.3.0}) for $\nu$ to obtain
\be
\nu=\frac{1}{2}\ln (R_+R_-)+\frac{1}{2} \int_0^\rho\Big(\frac{h_+}{R_+}
+ \frac{h_-}{R_-}\Big), \label{N3.8}
\ee
with $h_\pm$ given by (\ref{GP.3.01}) (we have $\ln (R_+R_-)|_{\rho=0}=0$
because of \eq{N3.5} and because $R_t|_{\rho=0}=0$ by regularity of the
metric on the symmetry axis). If the energy condition (\ref{NGP.5})
 holds,
(\ref{N3.6})--(\ref{N3.7.1}) and (\ref{N3.8}) imply
$$
\frac{h_+}{R_+}, \frac{h_-}{R_-}\in L^1 ([0,\infty)).
$$
{}From (\ref{GP.3.01}) it now follows:
\begin{Proposition}\label{PN3.1} Let $(\Si,\ga,K)$ be asymptotically flat
cylindrically symmetric inital data (in the sense of
(\ref{N3.6})--(\ref{N3.7.1}))
with matter satisfying the energy
condition (\ref{NGP.5}). Then we must necessarily have
\beq
&\sqrt{R}\umac _\pm,\quad  R^{-1/2} \asmalla _\pm\quad  \in\quad  L^2
([0,\infty)), \label{N3.8.0} \\
& R^{-1} \la^{ab} c_ac_b, \quad
RT_{\mu\nu}n^\mu(n^\mu\pm m^\mu)\quad  \in \quad L^1([0,\infty)). \nonumber
\eeq
\end{Proposition}

Under the hypothoses of
 asymptotic flatness we may without loss of generality assume,
rescaling $z$ if necessary, that
\be
\lim_{\rho\to\infty}\umac =0. \label{N3.14}
\ee
Moreover, (\ref{N3.11}) shows that we can redefine $\rho$ so that $R=\rho$ on
$\Si$. [Note, however, that we do {\em not} assume at this stage that this
will hold at later
times, with a metric of the form (\ref{GP.1})--(\ref{GP.1.1}). It will
nevertheless
be seen below that such a hypothesis could be assumed without loss of
generality for electro--vacuum metrics.] When \eq{N3.8.0} holds it can
be shown that $a=o(\rho)$ for large $\rho$, so that with those
normalizations it is easily seen that when $\rho$ tends to infinity
the geometry approaches   a flat conical geometry, with
opening angle equal to
\be
\anglevar _0=2\pi\exp\left\{-\frac{1}{2}\int_0^\infty\Big(\frac{h_+}{R_+} +
\frac{h_-}{R_-}
\Big)
d\rho\right\} \ .
\label{N3.15}
\ee
We thus obtain the well known ``positive energy theorem'' for cylindrically
symmetric initial data sets:

\begin{Proposition}\label{PN3.1.1} Under the hypotheses of
Proposition \ref{PN3.1}, the deficit angle
$$ \Delta \anglevar =2\pi-\anglevar _0=2\pi
\left(1-\exp\Big\{-\frac{1}{2}\int^\infty_0 \Big(\frac{h_+}{R_+} +
\frac{h_-}{R_-}\Big) d\rho\Big\}\right)
$$
satisfies
$$
0\leq\Delta\anglevar <2\pi.
$$
Moreover $\Delta\anglevar =0$ if and only if $(\Si,\ga,K)$ are initial data for
Minkowski space-time.
\end{Proposition}
A ($u$--dependent) quantity analogous to the opening angle $\theta_0$
defined above can also
be defined in the radiation regime\footnote{\label{scrifoot}Here and elsewhere,
when
talking about $\scri$ we mean the conformal boundary at future null infinity
of $2+1$ dimensional Minkowski space--time. Indeed from what is said
in this paper it follows that the coordinates in which the metric
takes the form  (\ref{GP.18})--(\ref{GP.19}) provide a natural
identification of $M/\R$, where $\R$ here refers to the orbits of the
Killing vector
$\partial/\partial z$, with $\R^{2,1}$ (this is of course {\em not} an
isometry).}
at $\scri$, {\em cf.\/} eq.\
\eq{angleatscri}; for vacuum metrics
this is briefly discussed at the end of Section \ref{global}.

Let us finally mention that the quantity
$$
\lim_{\rho\to \infty} g_{t\theta}
$$
is usually associated with  global rotation (``spinning strings", etc.),
whenever
it exists. Theorem
\ref{TPG.1} shows that this quantity  necessarily vanishes, when the
energy--momentum tensor satisfies the conditions of this Theorem and
when the asymptotic flatness condition (\ref{GP.23}) holds. In
particular there are no ``spinning'' purely vacuum ({\em
cf.\/} Theorem \ref{TPG.2}) or electrovacuum ({\em
cf.\/} Theorem \ref{TPG.3} below) asymptotically flat
space--times\footnote{We are grateful to P.\ Tod for pointing out this
implication to us.}.

\section{ Electrovacuum space-times.}
\label{electrovac}

In this Section we shall suppose that the metric satisfies the Einstein-Maxwell
equations. Let us start with a short discussion of the vacuum case.

\subsection{Reduced Vacuum Field Equations}
\label{vacuum}
Consider any cylindrically symmetric vacuum initial data $(\Si,\ga,K)$
such that (\ref{GP.23}) holds.  Theorem \ref{TPG.2} shows that we can
assume that the resulting space-time metric takes the form
(\ref{GP.18})--(\ref{GP.19}), with the Cauchy surface $\Si$ given by
an equation $t=i(\rho)$, for some smooth function $i:[0,\infty)\to\R$.
Parametrizing $\la_{ab}$ as in (\ref{GP.1.1}) we then have \be ds^2 =
e^{2(\nu-\umac )} (-dt^2+d\rho^2)+e^{2\umac } (dz+\asmalla d\anglevar
)^2+ \rho^2 e^{-2\umac } d\anglevar ^2. \label{3.1} \ee Equation
(\ref{Aequation}) specialized to vacuum implies that we can introduce
the Geroch--Ernst potential $\om$ \cite{Geroch,Ernst}, in terms of
which we have \beq
\part_t\asmalla  &=& -\rho e^{-4\umac }\om_{,\rho}, \label{3.2}\\
\part_\rho \asmalla  &=& -\rho e^{-4\umac }\om_{,t}. \label{3.2.1}
\eeq
The vacuum Einstein equations then yield a wave-map equation for a map
$\phi(t,\rho)=(\umac(t,\rho) ,\om(t,\rho))$ from $2+1$ dimensional
Minkowski space $(\R^{2,1}{},\eta)$,
$$
\R^{2,1}\approx\R^3,\quad \eta=-dt^2+d\rho^2+\rho^2 d\anglevar ^2,
$$
to the two-dimensional hyperbolic space $({}^2\cH,h)$:
\be
{}^2\cH\approx\R^2,\quad h_{ab}dx^adx^b=d\umac ^2+\frac{e^{-4\umac
}}{4} d\om^2.
\label{3.4}
\ee
Moreover, $\phi$ is invariant under rotations in the $t={\rm const}$
surfaces in
$\R^{2,1}$. The equations satisfied by $\phi$ are the variational equations for
the action
$$
I=\int_{\R^3} \eta^{\mu\nu} h_{ab} \frac{\part\phi^a}{\part x^\mu}
\frac{\part\phi^b}{\part x^\nu} dtdxdy,
$$
and we write them symbolically in the form
\be
\cD^\mu\phi_{,\mu}=0 \label{3.5}
\ee
Given a solution of (\ref{3.5}), $\asmalla $ can be obtained from (\ref{3.2.1})
using the regularity condition
\be
\asmalla (t,\rho=0)=0. \label{3.6}
\ee
$\nu$ is then given by (\ref{N3.8}), and we actually have
\be
\nu_\pm = \rho \left[\umac ^2_\pm +\frac{e^{4\umac }}{4\rho^2} \asmalla
^2_\pm\right]
= \rho \left[\umac ^2_\pm +\frac{e^{-4\umac }}{4} \om^2_\pm\right]\ .
 \label{3.7}
\ee
Let $f=\phi|_\Si$, $\dot f=\frac{\part\phi}{\part t}|_\Si$, be the Cauchy data
for
(\ref{3.5}). We can find non--trivial space--times 
with $(f,\dot f)$ --- 
compactly supported, by which we mean that $f$ is constant outside of a compact
set $\cC$ in $\R^2$, and $\dot f$ vanishes in $\R^3\setminus\cC$.
{}From the hyperbolic character of the semi-linear equation (\ref{3.5}) it
follows
that compactness of the support of $\phi$ is preserved by evolution,
consequently there exists a constant $C>0$ such that
\be
\frac{\part \umac }{\part x^\mu}\Big|_{\rho\geq C+|t|} = \frac{\part
\asmalla }{\part x^\mu} \Big|_{\rho\geq C+|t|} = \frac{\part\nu}{\part x^\mu}
\Big|_{\rho\geq C+|t|} =0. \label{3.9}
\ee
It follows that there exist constants $\umac _{i^\circ},\asmalla
_{i^\circ},\nu_{i^\circ}$
such that
\beq
\asmalla \Big|_{\rho\geq C+|t|}
 &=& \asmalla _{i^\circ}\equiv-\int^\infty_0 \Big(\rho e^{-4\umac }
\frac{\part\om}{\part t} \Big)\Big|_\Si d\rho \ ,
\label{3.11}
\\
\nu\Big|_{\rho\geq C+|t|}
&=& \nu_{i^\circ}\equiv\int^\infty_0\rho \Big[\umac _t^2 +
\umac ^2_\rho+\frac{e^{-4\umac }}{4} (\om_t^2+\om_\rho^2)\Big]\Big|_\Si
d\rho \ ,
\label{3.12}
\\
\umac\Big|_{\rho\geq C+|t|}
&=& \umac _{i^\circ}=0\ , \label{3.10}
\eeq
where the last equality in (\ref{3.10})  has been achieved by a rescaling of
$z$.
It follows that for $\rho\geq C+|t|$ the metric takes the form
\be
ds^2\Big|_{\rho\geq C+|t|}
= e^{2\nu_{i^\circ}} (-dt^2+d\rho^2)+
(dz+\asmalla _{i^\circ}d\anglevar )^2+\rho^2d\anglevar ^2, \label{3.14}
\ee
which is easily seen to be flat. This metric
is isometric to the one induced from the standard Minkowski metric
on the quotient of the set 
$M_C=\{x^\mu\in\R^4:x^2+y^2\geq C+|t|\}$
by the equivalence relation $\sim$,
where  $\sim$ is defined as
$$
(t,z,\rho,\anglevar )\sim(t,z+\asmalla _{i^\circ}\anglevar
_{i^\circ},\rho,\anglevar
+\anglevar _{i^\circ}) \ .
$$
Here
$\anglevar _{i^\circ}=\anglevar _0$ is
the opening angle given by (\ref{N3.15}),
\be
0<\anglevar _{i^\circ}=2\pi e^{-2\nu_{i^\circ}} \leq2\pi\ . \label{3.15}
\ee
The quantities (\ref{3.11}) and (\ref{3.12})
 have a direct interpretation in terms
of the wave-map equation: (\ref{3.12}) is, up to a factor $(2\pi)^{-1}$, the
(conserved) energy of the wave-map. On the other hand, $\asmalla _{i^\circ}$ is
a
conserved quantity for (\ref{3.5}) which is obtained as follows: Let
$Z^a\frac{\part}{\part y^a}$ be a Killing  vector for the metric $h_{ab}$ given
by (\ref{3.4}).  It is well known (and
in any case easily checked) that the quantity
\be
Q(Z,\Si) = \int_\Si \eta^{\mu\nu} h_{ab} \frac{\part\phi^a}{\part x^\mu}
Z^bd\Si_\mu 
\label{3.16}
\ee
(where $
 d\Si_\mu = \part_\mu\vee (dt\wedge dx\wedge dy)$, and $\vee$ denotes
contraction)
is (under appropriate asymptotic conditions
which are satisfied here)
independent of the choice of the asymptotically flat
Cauchy surface $\Si\subset \R^{2,1}$. Taking
$Z^a\frac{\part}{\part y^a} = \frac{\part}{\part\om}$ and
$\Si=\{t=0\}$ one finds
$$
Q(\frac{\part}{\part\om},\Si)=-2\pi \asmalla _{i^\circ}\ .
$$
Let us now describe how the above results generalize to the
electro--vacuum case:
\subsection{Electro--vacuum cylindrically symmetric initial data.}
Consider any globally hyperbolic electro--vacuum space--time $(\tilde M,g)$
with a Cauchy surface $\Sigma \approx \R^3$.
Let  $G=\R\times U(1)$ act
 on $\tilde M$ by isometries of $g$, the action on $\Sigma $ being that by
translations in $z$ and rotations  in the planes $z=\mbox{const}$.
Consider a source--free electromagnetic field  $F$ invariant w.r.t.
the Killing fields ${\partial/
\partial \theta}$, ${\partial/ \partial z}$, {\em i.e.\/}
$${\cal L}_{X_a}F=0, \quad a=1,2\ $$
(as before, ${\cal L}$ denotes a Lie derivative).
Let $x^0$ be any coordinate defined in a neighborhood of $\Sigma$ such
that $\Sigma=\{x^0=0\}$.
We can decompose $F$ into its electric and magnetic parts with respect to
the  surfaces $\{x^0=\mbox{const}\}$. Let $x^3$ be any coordinate on
$\Sigma$ which is constant on the orbits of the isometry group, extend
$x^3$ to a coordinate on some neighbordhood of $\Sigma$ in $\tilde M$ in any
way, preserving, however, the 
property of 
$x^3$ being invariant under $G$.
Now, the electric and magnetic vector densities  $\E^i$ and $\B^i$
obey ({\em cf. e.g.\/}
\cite{MTW}[Chap.21])
${\E^i}_{,i} ={\B^i}_{,i} = 0$ so, in the present case, we have
${\E^\rhos }_{,\rhos } = {\B^\rhos }_{,\rhos } = 0$.
Regularity at $x^\rhos  = 0$ implies $ \E^\rhos | _{x^\rhos  = 0} =
\B^\rhos | _{x^\rhos  = 0} = 0$ so that we get
\be
\label{EM.1}
\E^\rhos  = \B^\rhos  = 0\ .
\ee 
[Eq.\ (\ref{EM.1}) actually holds on that open connected subset $\tilde M'$
of $\tilde M$
which contains complete rays
 $\{(x^0,x^a,sx^3), s\in [0,1]\}$. 
Replacing $\tilde M$ by $\tilde M'$ we can
assume that  (\ref{EM.1}) holds on $\tilde M$.]
(\ref{EM.1}) shows that (\ref{GP.20})--(\ref{GP.21}) hold ({\em cf.\/}
also \cite{KT}), and, taking $(M,g)$ to be (perhaps an appropriate
subset\footnote{It will be seen in Section \ref{global}
that, under the conditions of  Theorem \ref{T4.2}, the
coordinates in which the metric takes the form
(\ref{GP.20})--(\ref{GP.21}) are global on the
maximal globally hyperbolic development of the data; {\em cf.\/}
Corollary \ref{C4.1}.})  the maximal globally
hyperbolic development of the data, from  Theorem \ref{TPG.1}
one obtains the following:
 \begin{Theorem}\label{TPG.3} Let $(\Si,\ga,K,A,E)$ be cylindrically symmetric
   initial data for electro--vacuum Einstein equations. Suppose
   moreover that there exists $C>0$ such that \be x^3\geq C : \quad
   Y^\mu\nabla_\mu R\Big|_\Sigma>0 \label{GP.23.new} \ee where $Y^\mu$
   is any {\em outwards directed} null vector orthogonal to the orbits of $G$,
   and $x^3\ge 0$ is any coordinate parametrizing the orbits of the
   group on $\Sigma$. Then there exists a globally hyperbolic
   electro--vacuum space-time $(M,g)$ with a metric of the form
   (\ref{GP.18})--(\ref{GP.19}) and an isometric embedding $i:\Si\to
   M$ such that $i(\Si)$ is a Cauchy surface for $(M,g)$.
\end{Theorem}

\subsection{The electromagnetic gauge conditions}

In the coordinate system in which
(\ref{GP.18}) holds the hypersurface $i(\Sigma)$, where $i$ is given
by Theorem \ref{TPG.3}, is a graph of a function $t=f(\rho)$.
Define new coordinates $x^\mu$ by $x^0=t-f(\rho)$, $x^3=\rho$; note
that in the coordinate system $(x^0,x^a,x^3)$ as defined here the
metric does {\em not} take the
form (\ref{GP.18}), but it is {\em block--diagonal} with respect
to the pairs of variables $(x^0,x^3)$ and $x^a$, $a=1,2$.
The equation
$dF = 0$ on the simply connected manifold  $M$ 
together with the Poincar\'e Lemma show that
$ F = dA$.
We can always impose the  {\it gauge condition}  $A_0 = 0$.
 The equations $
\E^\rhos
= 0 = A_0 $ imply $
A_{\rhos , 0} = 0$.
Moreover
the equation
$\B^\rhos  = 0$ yields
$A_{\theta, z}  - A_{z, \theta} = 0$.
Let
$$ s = - \int^\rho_0  A_\rho(\rho', z, \theta)d\rho'\ .
$$
As $A_\rho$ is independent of $t$, 
the gauge transformation $A_\mu\to A_\mu+s_{,\mu}$
leaves $A_0 = 0$ and transforms $A_\rho$ to zero.
In the new gauge we have
$$A_0 \equiv 0,\quad A_\rho \equiv 0\ ,
$$
 so that
$$F_{\rho\theta} = -A_{\theta, \rho} ,\quad F_{t\theta} = -
A_{\theta, t} ,\quad F_{\rho z} = -A_{z, \rho}  ,\quad F_{tz} = -
A_{z, t}
\ ,
$$
$$F_{t\rho}   \equiv 0, \quad F_{\theta z} = 0 =
\partial_\theta A_z - \partial_zA_\theta\ .
$$
On the initial surface $\Sigma$ we choose
$$
A_\theta\Big| _\Sigma = - \int^\rho_0 F_{\rho\theta}\Big| _\Sigma
d\rho'
, \qquad
A_z\Big| _\Sigma = - \int^\rho_0 F_{\rho z}\Big| _\Sigma
d\rho' \ ,
$$
$$A_{\theta, t}\Big| _\Sigma = - F_{t\theta}
, \qquad
A_{z,t}\Big| _\Sigma = - F_{tz}\ ,$$
which yields $\partial_\theta A_\theta| _\Sigma =
\partial_\theta A_z| _\Sigma = \partial_z A_\theta| _\Sigma
=  \partial_z A_z| _\Sigma = 0$ and thus $(\partial_\theta A_z -
\partial_z A_\theta )| _\Sigma = F_{\theta z}| _\Sigma = 0$.
However, since in the $A_0 = A_\rho = 0$ gauge we have $A_{\theta, t} =
- F_{t\theta}$, $A_{z,t} = -F_{tz}$ it is clear that $A_\theta,
A_z$ remain invariant with respect to ${\partial/\partial\theta}$,
${\partial/ \partial z}$ off the initial surface.  Thus we may
assume without loss of generality that
\be
\label{EM.2}
A = A_\theta  d\theta + A_z dz\ ,
\,
\ee
where
\be
\label{EM.3}
\partial_\theta A_\mu = \partial_z A_\mu = 0\ .
\ee
Note that (\ref{EM.2})--(\ref{EM.3}) are invariant under changes of
coordinates
in the $(t,\rho)$ plane. We can thus go to the coordinates in which
the metric takes the form (\ref{GP.18}), with
(\ref{EM.2})--(\ref{EM.3}) still holding.

\subsection{Reduced Electro--Vacuum Field Equations}
\label{reducedelvac}

It is convenient to introduce the gravitational and electromagnetic twist
potentials $\{\omega , \eta\}$
satisfying \cite{Ernst,Kinnersley,Mazur,VM1,VM2}:
 $$a_{, t} = -\rho e^{-4\gamma}(\omega_{,\rho} + A_z \eta_{,\rho})\ ,
\qquad
a,_\rho = -\rho e^{-4\gamma}( \omega_{,t} + A_z \eta_{,t})\ ,
$$
$$
 A_{\theta, \rho}-aA_{z,\rho}
= \rho e^{-2\gamma} \eta_{,t}\ ,\qquad
A_{\theta,t} - aA_{z,t} = \rho
e^{-2\gamma} \eta_{,\rho}
\ .
$$
This corresponds to solving the 2 divergence constraints
given in Eq. (2.6) of \cite{VM1}, 
{\em i.e.},
${\tilde f^a}{}_{,a} = 0$, $ {\E^a}_{,a} = 0$ by $\tilde f^a =
\epsilon^{ab}\omega,_b$,
$ \E^a = \epsilon^{ab}\eta_{,b}$.
Now, define  $\lambda:= A_z$ and consider the map,  taking values in
$\R^4$, defined by  $\{\gamma , \omega, \lambda, \eta\}$.
One can show, after introducing $\eta$ as in \cite{VM1} (setting to
$0$ the Higgs field $\Phi$ of \cite{VM1}) that the electro--vacuum
field equations yield a
wave-map for a map $\phi = \{\gamma, \omega, \lambda, \eta\}$ from $2 + 1$
 dimensional Minkowski space $(\R^{2,1}{}, \eta) $
to the $4$--dimensional Riemannian manifold $(\R^4, h)$:
\be
\label{above}
h_{ab}dx^adx^b= 4(d\gamma)^2 + e ^{-2\gamma}\left[(d\lambda)^2 +
(d\eta)^2\right]
 + e^{-4\gamma}(d\omega + \lambda d\eta)^2\ .
\ee
Moreover $\phi$ is invariant under rotations generated by $\partial/
\partial\theta$ in $\R^{2,1}$.  The equations satisfied by $\phi$ are the
variational equations for the action
$$ I = \int_{\R^3} \eta^{\nu\mu}h_{ab}\frac{\partial\phi^a}{\partial
x^\nu}\frac{\partial\phi^b}{ \partial x^\mu}\sqrt{-\det\eta} \,
dt\,dx\,dy \,
$$
and we write them symbolically in the form
$$
{\cal{D}}^\mu\phi_{,\mu} = 0 \ .
$$
Given a solution, we can recover $a$ and $A_\theta$ from the equations above
defining the twist potentials
(upon imposing regularity $a(t, \rho = 0) =
A_\theta (t, \rho = 0) = 0$). Note that $A_z = \lambda$ so the only remaining
unknown is the metric function $\nu$ which is determined from
$$\nu_{, \rho} \pm \nu_{,t} = \frac{\rho}{ 4} h_{ab}
({\Xphi^a}_{,\rho} \pm {\Xphi^a}_{,t})
({\Xphi^b}_{,\rho} \pm {\Xphi^b}_{,t})\ ,$$
where $\{\Xphi^a\} = \{\gamma, \omega, \lambda, \eta\}$ and $h_{ab}$ is given
above, eq.\ (\ref{above}),
{\em i.e.},
$$
\nu_\pm = \rho\Big\{(\gamma_\pm) ^2 +\frac{e^{-2\gamma}}{
4}\left[(\lambda_\pm)^2 +
(\eta_\pm)^2 \right]+ \frac{e^{-4\gamma}}{4}(\omega_\pm + \lambda
\eta_\pm)^2\Big\}\ .
$$
The discussion after eq.\ (\ref{3.7}) 
goes through as before with eq.
(\ref{3.11}) modified to read
$$
a\Big| _{\rho \geq C+|t|} = a_{i^o} \equiv - \int^\infty_0 \rho
e^{-4\gamma}(\omega_{,t} + \lambda \eta_{,t})\Big| _\Sigma \,d\rho\ ,
$$ 
and
$$
\nu\Big| _{\rho\geq C+|t|} = \nu_{i^o}\equiv \int^\infty_0 \rho
\Big\{(\gamma_{,\rho})^2 +\frac{e^{-2\gamma}}{ 4}\left[(\lambda_{,\rho})^2
+ 
(\eta_{,\rho})^2 \right]
+\frac{e^{-4\gamma}}{ 4}
(\omega_{,\rho} + \lambda\eta_{,\rho})^2\qquad
$$
$$
\hspace{2.5cm}+
(\gamma_{,t})^2 + \frac{e^{-2\gamma}}{ 4}\left[(\lambda_{,t})^2 +
(\eta_{,t})^2\right]+\frac{e^{-4\gamma}}{ 4}
(\omega_{,t} + \lambda
\eta_{,t})^2\Big\}\  d\rho\ .
$$ Note that again up to a constant factor $\nu_{i^o}$ is the
conserved energy of the wave map; similarly, $a_{i^o}$ is again
conserved as because $\partial/ \partial\omega$ is a Killing vector
field of the metric $h_{ab}$ ({\em cf.\/} the discussion around eq.\
(\ref{3.16})).  In fact, there are a total of 8 conserved quantities
as defined by (\ref{3.16}) corresponding to the $SU(2,1)$ isometry
group of the metric $h_{ab}$ ({\em cf.\/}
\cite{Kinnersley,Mazur,VM2}). A subset of these Killing fields given
by
$$
Z_{(1)} = \frac{\partial}{ \partial\omega} ,\quad Z_{(2)} = \frac{\partial}{
\partial\eta} ,\quad Z_{(3)} = \frac{\partial}{\partial\lambda} - \eta
\frac{\partial}{\partial\omega} ,
$$
$$
Z_{(4)} = (\frac{\partial}{\partial\gamma} + \eta \frac{\partial}{
\partial\eta} +  \lambda \frac{\partial}{ \partial\lambda} + 2\omega
\frac{\partial}{ \partial\omega})
$$
defines a basis to the tangent space at every point of $(\R^4, h)$ and in
fact generates the action of a transitive subgroup of $SU (2,1)$.  One sees
this for example by integrating the flow generated by  $Z = \sum^4_{i=1}
\alpha^i Z_{(i)}$, $\alpha^i $   constant and showing that the resulting action
is transitive explicitly.  That, in any case $SU (2,1)$ acts transitively on
$(\R^4, h)$ follows from its structure as a coset space ({\em cf.\/}
\cite{Mazur,Kinnersley}).  The Lie subalgebra generated by the $Z_{(i)}$'s is
$$[Z_{(1)} , Z_{(2)}] = [Z_{(1)} , Z_{(3)}] = 0\ ,$$
$$[Z_{(1)} , Z_{(4)}] = 2Z_{(1)} ,\quad [Z_{(2)} , Z_{(3)}] = -
Z_{(1)}\ ,$$
$$[Z_{(2)} , Z_{(4)}] = Z_{(2)} , \quad[Z_{(3)} , Z_{(4)}] = Z_{(3)}\ .$$
The metric $h$ admits an orthonormal frame $\{X_{(i)}\}$ which
determines a Lie algebra isomorphic to that of the $Z_{(i)}$'s.
Defining
$$X_{(1)} = \frac{1}{ 2} \frac{\partial}{ \partial\gamma} , \quad
X_{(2)} = e^\gamma
\frac{\partial}{ \partial\lambda}\ ,$$
$$X_{(3)} = e^{2\gamma} \frac{\partial}{ \partial\omega} ,\quad  X_{(4)} =
e^\gamma\frac{\partial}{\partial\eta} - \lambda
e^\gamma\frac{\partial}{\partial\omega}\ ,$$
one finds that $h(X_{(i)}, X_{(j)}) = \delta_{ij}$ and that
$$
[X_{(1)}, X_{(2)}] = \frac{1}{2} X_{(2)} ,\quad  [X_{(1)}, X_{(3)}] =
X_{(3)},\quad [X_{(1)}, X_{(4)}] = \frac{1}{2} X_{(4)}\ ,
$$
$$
 [X_{(2)}, X_{(3)}] = 0,\quad [X_{(2)}, X_{(4)}] = - X_{(3)} ,\quad
[X_{(3)}, X_{(4)}] = 0\ .$$
The isomorphism of Lie algebras is seen by making the correspondence $X_{(1)}
\to - Z_{(4)}/{  2}$, $X_{(2)} \to Z_{(2)}$, $ X_{(3)} \to Z_{(1)}$ and
$X_{(4)} \to Z_{(3)}$.  Note however, that the group action generated by the
$X_{(i)}$'s is not in general an isometry of $h$.  As will be seen
in the next section, the existence of an
orthonormal frame with bounded structure functions (in this case constants)
plays a key note in the 
global existence theorem of Christodoulou and Tahvildar--Zadeh.

Another key element in the application of the Christodoulou --
Tahvildar--Zadeh theorem involves verifying
that for any point $p$ in the manifold $(\R^4, h)$ the geodesic sphere
$\Sigma(p,s)$ of radius $s$ centered at $p$ has a second fundamental form
$k_{ab}$ whose eigenvalues obey certain bounds.  Since $(\R^4, h)$ is a
homogeneous space ({\em cf.\/} \cite{Mazur} or the explicit argument above) it
suffices to verify the conditions on $\Sigma(p,s)$ for any fixed point $p$.
Introducing two complex coordinates $\{w^1, w^2\}$ via the definitions
$$
\frac{1-w^1}{ 1 + w^1} = e^{2\gamma} + \frac{1}{ 4}(\eta^2 - \lambda^2) +
i(\omega + \frac{1}{ 2}\eta\lambda) \ ,
$$
$$
\frac{w^2}{ 1+w^1} =
\frac{1}{ 2} (\eta + i\lambda)\ ,
$$
one finds (\`a la Mazur) that the metric $h_{ab}$ becomes
$$dh^2 = \frac{4}{ (1-|  w^1| ^2-|  w^2| ^2)^2}
\Big\{dw^{1*}dw^1 + dw^{2*}dw^2 \qquad\qquad
$$
$$
\hspace{2cm}
- (w^2 dw^1 - w^1dw^2)(w^{2*}dw^{1*}  - w^{1*}dw^{2*})\Big\}\ .
$$
Taking $w^1 = x + iy$, $w^2 = u + iv$ and introducing the ``spherical"
coordinates $\{r,  \theta,,\varphi, \chi\}$ defined by
$$
x = r \cos(\varphi) \sin (\chi) \sin(\theta),\quad
y = r \sin (\varphi) \sin(\chi) \sin(\theta)\ ,
$$
$$ u = r \sin(\chi) \cos(\theta),\qquad v = r \cos (\chi)\ ,$$ and
reexpressing $h_{ab}$, once again we obtain a convenient coordinate
system for the computation of the extrinsic curvature of the geodesic
spheres centered at the origin $(r = 0)$ of the new coordinate system.
The result of a lengthy calculation (done using {\sc Mathematica})
shows that the $3$ eigenvalues of $k_{ab}$ are in fact independent of
direction and given explicitly by
$$k_1 =
\coth({s})
,\quad k_2 = k_3 =
{1\over 2} \coth ({s\over 2})\ .$$
Note that there exists constants $c, C > 0$ such that the smallest eigenvalue
$\lambda (= k_2 = k_3)$ satisfies
$ s\lambda \geq c$
and such that the largest eigenvalue  $\Lambda ( = k_1)$ satisfies
$s\Lambda \leq C (1 + s)$.

\section{Global existence, strong cosmic censorship.}
\label{global}

In the previous sections we have seen how to reduce the dynamics of
cylindrically symmetric electro--vacuum space--times to a wave-map
problem. Global existence and
asymptotic properties of this last problem have been studied recently by
D.~Christodoulou and A.S.~Tahvildar-Zadeh \cite{ChTZ1,ChTZ2}. These
authors consider rotation-invariant maps $\phi:(\R^{2,1}, \eta)\to(N,h)$, where
$(N,h)$ is a complete Riemannian manifold. Passing to the universal cover of
$N$ if necessary we may assume that $N$ is simply connected. Christodoulou and
Tahvildar-Zadeh moreover assume that

\begin{enumerate}
\item[C1.]
\label{C1} There exists an orthonormal frame of vector fields $\Om_A$ on $N$
such that the functions $e^A_{BC}$ defined by
\be
[\Om_B,\Om_C]=\sum_A e^A_{BC}\Om_A \label{4.1}
\ee
are uniformly bounded on $N;$

\item[C2.]
\label{C2} For each $p\in N$ let $\Si(p,s)$ denote the geodesic sphere of
radius $s$ centered at $p$, and let $k_{AB}$ be its extrinsic curvature. There
exists a constant $c>0$ such that the smallest  eigenvalue $\la$,
respectively the largest eigenvalue  $\Lambda$,
of $k_{AB}$ (with respect to $h$),
satisfies
\be
s\la\geq c^{-1},\quad s\Lambda\leq c(1+s) \ .
\label{4.2}
\ee
\end{enumerate}

Let us mention that (\ref{4.2}) implies that the geodesic spheres
are
differentiable spheres (recall that we have excluded cut points by assuming
that $N$ is simply connected). It follows that $N$ is topologically trivial.
Let us also mention, that the Eschenburg comparison theorem \cite{Eschenburg}
[Theorem 3.2] shows that C2
 will hold when the sectional curvatures $\ka$
of $(N,h)$ satisfy
$$
-C\leq\ka\leq 0\ ,
$$ for some constant $C\geq0$, and that this criterion is satisfied by
the target space which occurs in the electro--vacuum case
\cite{Mazurreview}.

Under 
C1 and 
C2 we have the following (by definition, $\phi$ is
rotation invariant if $\phi$ depends only upon $\rho$ and $t$):

\bt\label{T4.1} {\rm \cite{ChTZ1,ChTZ2}}: Let $\Si$ be a rotation-invariant
Cauchy
surface in $\R^{2,1}$ and let $\phi_0,\dot\phi_0$ be any smooth
rotation-invariant Cauchy data for the wave map equation for a map
$\phi:(\R^{2,1},\eta)\to(N,h)$, where $(N,h)$ satisfies the hypotheses
C1 and C2. Then

1. There exists a unique smooth-map $\phi:\R^{2,1}\to N$ satisfying the
wave-map equation and assuming the Cauchy data $\phi_0,\dot\phi_0;$

2. Suppose moreover that $\dot\phi_0$ has support in a compact set
$\cC\subset\Si$, while
$\phi_0$ maps $\Si\setminus\cC$ into a point.
For $t\geq0$ we have the following pointwise estimates for the derivatives of
$\phi$:
\beq
\|\part_-\phi\|\equiv\|(\part_\rho-\part_t)\phi\| &\leq&
\frac{C}{(1+t+\rho)^{1/2}(1+|t-\rho|)}, \label{4.4}\\
\|\part_+\phi\|\equiv\|(\part_\rho+\part_t)\phi\| &\leq&
\frac{C}{(1+t+\rho)^{3/2}}, \label{4.5}\\
\|\partial_\mu\partial_\nu \phi\| &\leq& C, \label{4.5.0}
\eeq
for some constant $C$.
Moreover, if we denote by 
$C^+_u$  the interior of the
future light cone with vertex at $(t=u,\rho=0)$, then  there exists a constant
$C_1\geq0$ such that for $t\geq\rho$ we have
\be
{\rm diam\/} (\phi(C^+_{t-\rho})) \leq \frac{C_1}{(1+|t-\rho|)^{1/2}},
\label{4.3}
\ee
Here the norm $\|\part_\pm\phi\|$ is taken with respect to the metric $h$ on
$N$, and ${\rm diam}(\Om)$ denotes the diameter of a set $\Om$.
\et
[The second derivatives estimates (\ref{4.5.0}) are proved in Section
4 of \cite{ChTZ2}. We have been informed by D.\
Christodoulou\footnote{D.\ Christodoulou, private communication.} that
 under the hypotheses above for $t\ge 0$ one can actually prove the
following decay estimates
\be
\label{christodoulouestimate}
\forall\  n,m\in {\N}\cup\{0\}, \ n+m>0:\qquad \|\partial^n_+\partial^m_-
\phi\|
\leq C (1+t+\rho)^{-1/2-n}(1+|t-\rho|)^{-m}\ ,
\ee
for some constant C.]

{}For $t\leq0$
estimates analogous to (\ref{4.4})--(\ref{4.3}) 
immediately follow from (\ref{4.4})--(\ref{4.3}) and from
time-reversal invariance of the wave-map equation.

Let us first consider the vacuum Beck \cite{Beck} space--times
(``polarized'' metrics), $\asmalla \equiv0$.  Eq.~(\ref{3.5}) reduces
to a linear scalar wave-equation on 3-dimensional Minkowski
space-time, so that the global existence immediately follows.
Moreover the asymptotic estimates (\ref{4.4})--(\ref{4.5}) readily
follow from the explicit representation of solutions of the scalar
wave--equation on 3-dimensional Minkowski space-time; an exhaustive
analysis of this case can be found in \cite{ABS}.

As discussed in Section \ref{reducedelvac}, the hypotheses of
Theorem~\ref{T4.1} are satisfied by the target manifold $(N,h)$ of the
wave-map associated with the electro--vacuum Einstein equations, so
that Theorem~\ref{T4.1} implies global existence on $\R^4$ of
cylindrically symmetric electro--vacuum space-times. The main result
of this paper is the following:

\bt \label{T4.2} Let $(\Si,\ga,K,A,E)$ be smooth cylindrically
symmetric Cauchy data for electro--vacuum Einstein equations such that
the initial data for the corresponding wave--map are supported in some
compact set $\cC\subset \Si/\R$ . Then the maximal globally hyperbolic
development $(M,g)$ of $(\Si,\ga,K,A,E)$ is causally geodesically
complete, hence inextendible. In particular, Strong Cosmic Censorship
holds in this class of space-times.  \et

{\bf Proof}: The proof of Theorem \ref{T4.2} consists in obtaining
estimates on the geometry of $(M,g)$ which will allow us to show
causal geodesic completeness of
$(M,g)$. By 
Theorems \ref{TPG.1} and 
\ref{T4.1} it is sufficient to
consider metrics on $\R^4$ of the form (\ref{GP.18})--(\ref{GP.19}).
 We shall prove
the result in vacuum, the analysis of the electro--vacuum case is essentially a
repetition of the arguments below.
 Replacing $t$ by $t+T$ if
necessary, for some appropriately chosen constant $T$, we may without loss of
generality assume that
$\cC\subset C^+_0\cap\Si$. By time--reversal invariance of the equations it is
sufficient to prove causal
future geodesic completeness, and henceforth we shall only consider
$M^+=\{p\in\R^4:\ t(p)\geq0\}$. It follows that
(\ref{3.11})--(\ref{3.10}) hold in
$M^+$ with $C=0$, and to understand the geometry of $(M,g)$ we 
will have to analyze
the behavior of $\asmalla $ and $\nu$ in the set ${\{t\geq\rho, t\ge
T\}}$. By (\ref{4.4})--(\ref{4.3}) there exist
constants $C$, $\umac _{i^+}$ and $\om_{i^+}$ such that we have
\beq
|\part_+\umac |+ |\part_+\om|&\leq& \frac{C}{(1+t+\rho)^{3/2}}\ ,
\label{4.6.1}\\
|\part_-\umac |+ |\part_-\om|& \leq& \frac{C}{(1+|t-\rho|)(1+t+\rho)^{1/2}} \
,
\label{4.6.2}
\\
|\umac -\umac
_{i^+}|+|\om-\om_{i^+}|&\leq&\frac{C}{(1+|t-\rho|)^{1/2}}\ ,\label{4.6} \\
|\part_{\mu}\part_{\nu}\umac |+ |\part_{\mu}\part_{\nu}\om|& \leq& C \
{}.
\label{4.6.3}
\eeq
In what follows the letter $C$ denotes a positive constant the value of which
may vary from line to line. From (\ref{3.2})--(\ref{3.2.1}), (\ref{3.7}) and
(\ref{4.6.1})--(\ref{4.6}) it follows that we have the estimates (recall that
$\part_\pm=\part_\rho\pm\part_t$)
\beq
|\part_+\asmalla | &\leq& \frac{C\rho}{(1+t+\rho)^{3/2}}\ , \label{4.7}\\
|\part_-\asmalla | &\leq& \frac{C\rho}{(1+|t-\rho|)(1+t+\rho)^{1/2}} \
,
\label{4.8}\\
|\part_+\nu| &\leq& \frac{C\rho}{(1+t+\rho)^3} \ ,\label{4.9}\\
|\part_-\nu| &\leq& \frac{C\rho}{(1+|t-\rho|)^2(1+t+\rho)}\ . \label{4.10}
\eeq
By direct integration of (\ref{4.9}) along the null geodesics
$u=\mbox{\rm const}$ one
finds that the function
\be
\nu_{\scri^+}(u)\equiv\lim_{\rho\to\infty}\nu(t=u+\rho,\rho) \label{4.11}
\ee
exists, and we have
\beq
|\nu_{\scri^+}(u)| &\leq& \frac{C}{1+u}, \label{4.12}\\
|\nu(t,\rho)| &\leq& C \left(\frac{1}{1+|t-\rho|} +
\frac{1+t+3\rho}{(1+t+\rho)^2}\right);\ \label{4.13}
\eeq
in particular $\nu$ is uniformly bounded on $M^+$.
{}From eq.\ (\ref{4.7}) we also  find
\beq
|\frac{\asmalla }{\rho}| &\leq& \frac{1}{\rho}\int^\rho_0 |\part_+\asmalla
|(
t-\rho+s,s)ds
\nonumber\\
&\leq& C \left(\frac{2}{(1+t+\rho)^{1/2}+(1+t-\rho)^{1/2}} -
\frac{1}{(1+t+\rho)^{1/2}}\right). \label{4.14}
\eeq
(\ref{4.14}) shows that $|{\asmalla }/{\rho}|$ is uniformly bounded on $M^+$,
and
uniformly decays to zero as $t\to\infty$.
It is now easy to check that
for $\rho\ge 1$  the inequalities \eq{A.4} of Appendix \ref{aproof}
are satisfied. By rotational symmetry the derivatives $\partial
 \om/\partial\rho $ and $\partial
 \umac/\partial\rho $ vanish at $\rho =0$, and from the second
 derivative estimate \eq{4.6.3} it follows that
 $$
 \rho\le 1:\qquad |\partial_\rho\umac | +|\partial_\rho\om |\le C
 \rho\ .
 $$
 It is easily seen now that \eq{A.4} holds for all $\rho$, and our
 claims follow by  Proposition \ref{PA.1} of the
 Appendix \ref{aproof}.\hfill $\Box$

 From Theorems \ref{TPG.3}, \ref{T4.1} and \ref{T4.2} we obtain the
 following:

\begin{Corollary}
\label{C4.1}
For initial data as in Theorem  \ref{T4.2}, the corresponding maximal
globally hyperbolic development 
is isometrically diffeomorphic  to $\R^4$ 
with a metric of the form
(\ref{GP.18})--(\ref{GP.19}).
\end{Corollary}

It is of some interest to enquire about completeness of affinely
parametized trajectories of charged test particles. Using estimates
similar to the ones given in the proof of Theorem \ref{T4.2} one can
check that the inequalities (\ref{D.5}) are satisfied; in particular
one finds that there exists a constant $C$ such that
$$
|\nu|+\Big|\frac{a}{\rho}\Big|+\Big|\frac{A_\theta}{\rho}\Big|\le C \ .
$$
[Actually the somewhat sharper estimates for $\nu$ and $a$ given in
the proof above still hold; moreover $A_\theta/\rho$ satisfies an
estimate of the form (\ref{4.14}).] It then follows
from the arguments of Appendix \ref{motioncharged} that for all
space--times as considered in Theorem \ref{T4.2} the trajectories of
non--tachyonic charged test particles are affinely complete.

 Let us finally note that  the angle
 \be
\label{angleatscri}
\theta(u)=2\pi e^{-\nu_{\scri^+}(u)} \ee can be thought of as the
``instantaneous conical angle\footnote{Here we are referring to the
  fact that the space--time geometry near the $2+1$ dimensional Scri
  ({\em cf.\/} footnote \ref{scrifoot}) ``looks conical''; no
  singularity of the geometry of Scri is implied.} of the geometry at
$\{u=\mbox{\rm const}\}\cap\scri$" ({\em cf.\/} the discussion at the
end of Section \ref{cylindrical}). In particular eq.~(\ref{4.12})
gives a bound on the rate at which $\theta(u)$ approaches $2\pi$ as
$u$ tends
to infinity.  
Now under the hypotheses of
Theorem \ref{T4.2} and assuming vacuum field equations for simplicity,
it is an easy exercice to show from 
the estimates \eq{christodoulouestimate} that the limit
$$ \lim_{\rho\to\infty}\rho \left[\umac ^2_- +\frac{e^{-4\umac }}{4}
\om^2_-\right](u+\rho,\rho)
$$ exists, and is a smooth function of $u$. This and \eq{3.7} imply
then that $\nu_{\scri^+}(u)$ is a smooth, monotonously decreasing
function of $u$; indeed the limit as $\rho\to\infty$, $t=u+\rho$,
$u=\mbox{\rm const}$, of equation 
\eq{3.7} gives a Fock--Trautman--Bondi--Sachs type mass--loss formula
\cite{BMS,Fock,Sachs,Trautman} for cylindrically symmetric waves.

%
%
%
%
%
%
%
\appendix
%
%
%

 \section{Causal geodesic completeness of some metrics}
 \label{aproof}


 In this Appendix we shall prove causal geodesic completeness of a family of
 metrics on $M=\R^4$ with a cylindrically symmetric metric of the form
 \beq
 ds^2 &=& h_{AB}dx^Adx^B+ \la_{ab}(dx^a+M^a_Adx^A) (dx^b+M^b_Bdx^B)\ ,
\label{A.1}\\
 h_{AB} &=& e^{2\psi}\eta_{AB},\qquad
  \eta_{AB}=\mbox{diag}(-1,1)\ , \label{A.2}
 \eeq
 and all the functions above depend only upon $x^A,\ A=0,1$. [It should be
clear
 from the arguments below that the proof goes through for more general
 topologies and actions of the symmetry group, under suitably modified
 assumptions. 
 More precisely, the arguments below apply to 
 $m+2$ dimensional
 space-times $(M,g)$ 
 with $m$ Killing vectors.] For the electro--vacuum
 space-times it would have been sufficient to assume
 $M^a_A=0$, $\det\la_{ab} 
 =\rho^2$. These assumptions, however, are
 unnecessary in the argument below.

 To avoid ambiguities, we shall say that  a metric of the form (\ref{A.1}) is
of
 differentiability class $C^k$ if the 
 components of the metric tensor in
 coordinates $(t,x,y,z)$ defined by
 $$
 t,\ x=\rho\cos\anglevar ,\ y=\rho\sin\anglevar ,\ z
 $$
 are, locally, of $C^k$ differentiality class (no uniform bounds are implied).
 We have the following:

 \begin{Proposition}\label{PA.1} Let $M=\R^4$, consider a cylindrically
 symmetric $C^2$ metric of the form (\ref{A.1})--(\ref{A.2}), with
 \be
 \la_{ab}dx^adx^b= e^{2\umac }(dz+\asmalla d\anglevar )^2+R^2 e^{-2\umac
}d\anglevar ^2. \label{A.3}
 \ee
 Suppose that there exists a constant $C$ such that
 \beq
 &&|\umac |+|\frac{\part \umac }{\part t}|+|\psi|+|\frac{\part\psi}{\part t}| +
  \frac{1}{R^2}\Big|\frac{\part \umac }{\part t} +
 \frac{\part\psi}{\part t} - \frac{1}{R}
   \frac{\part R}{\part t}\Big| (1+
 |\asmalla |^2)\qquad\qquad\qquad
 \nonumber
 \\
 &&\qquad\qquad\qquad 
 +  \frac{1}{R^2}\Big|\frac{\part \asmalla }{\part t}\Big| (1+|\asmalla |) +
 \Big|\frac{\part M^a_A}{\part t} - \frac{\part M^a_t}{\part x^A}\Big|
  +|M^z_A| \leq C\ .
 \label{A.4}
 \eeq
 Then $(M,g)$ is globally hyperbolic, the level sets of $x^0=t$ are Cauchy
 surfaces, and $(M,g)$ is causally geodesically complete.
 \end{Proposition}

 {\bf Remark:}
 Let us mention that the regularity of the metric at the axis $\rho=R=0$
implies,
 among others,
 $$
 \lim_{\rho\to0} \left(\frac{\part \umac }{\part t} + \frac{\part\psi}{\part t}
-
 \frac{1}{R} \frac{\part R}{\part t}\right) =0.
 $$

 {\bf Proof}: The equations for affinely parametrized geodesics are
 easily found from the variational principle for the action
 \be
 I=\int \left\{h_{AB} \dot x^A\dot x^B + \la_{ab} (\dot x^a + M^a_A\dot x^A)
 (\dot x^b+M^b_B\dot x^B)\right\} ds, \label{A.5}
 \ee
 where a dot over a quantity denotes a derivative with respect to the affine
 parameter $s$. The variation of (\ref{A.5}) with respect to $\dot x^a$ gives
the
 conservation equations $dp_a/ds=0$, where
 \be
 p_a = \la_{ab} (\dot x^b+M^b_A\dot x^A)\quad \Rightarrow\quad
  \dot x^a=\la^{ab} p_b-M^a_A \dot x^A,
 \label{A.6}
\ee
and where $ \la^{ab}$ is the matrix inverse to $ \la_{ab}$.
 (\ref{A.6}) and the affine parametrization condition give
 \be
 (\frac{dt}{ds})^2 = (\frac{d\rho}{ds})^2 + e^{-2\psi} (\epsilon+\la^{ab}
 p_ap_b), \label{A.7}
 \ee
 with $\epsilon=0$ for null geodesics and $\epsilon=1$ for timelike ones.
 Varying (\ref{A.5}) with respect to $\dot x^A$ one obtains
 \beq
 \frac{d}{ds} \left(e^{2\psi}\eta_{AB}\dot x^A\right) &=& -\frac{1}{2}
 \frac{\part\la^{ab}}{\part x^B} p_ap_b - (\epsilon+\la^{ab}p_ap_b)
 \frac{\part\psi}{\part x^B} \nonumber\\
 &+& p_a\left(\frac{\part M^a_A}{\part x^B} - \frac{\part M^a_B}{\part
x^A} \right)\dot x^A.
 \label{A.8}
 \eeq
 We have
 \be
 \la^{ab}p_ap_b=\frac{e^{2\umac }}{R^2} (p_\anglevar  -\asmalla
p_z)^2+e^{-2\umac }p_z^2, \label{A.9}
 \ee
 so that (\ref{A.8}) with $B=0$ gives
 \beq
 \frac{d}{ds} \left(e^{2\psi}\frac{dt}{ds}\right) &=&
\frac{e^{2\umac}}{R^2}
\left(\frac{\part \umac }{\part
 t} + \frac{\part\psi}{\part t} - \frac{1}{R} \frac{\part R}{\part t}\right)
 (p_\anglevar -\asmalla p_z)^2 \nonumber\\
  && \quad -\frac{e^{2\umac }}{R^2} \frac{\part \asmalla }{\part t}
 (p_\anglevar -\asmalla p_z) + e^{-2\umac }
 \left(\frac{\part\psi}{\part t} - \frac{\part \umac }{\part t}\right)
 p_z^2 \nonumber\\
 &&\quad +\epsilon
 \frac{\part\psi}{\part t}+
  p_a \left(\frac{\part M^a_0}{\part x^A} - \frac{\part M^a_A}{\part t}\right)
\dot x^A.
 \label{A.10}
 \eeq
 (\ref{A.7}) gives $|\frac{d\rho}{ds}|\leq|\frac{dt}{ds}|$, and assuming
 $\frac{dt}{ds}>0$ from (\ref{A.4}) and (\ref{A.10}) one obtains
 \be
 \frac{d}{ds}(e^{2\psi}\frac{dt}{ds})\leq C(1+e^{2\psi}\frac{dt}{ds}),
 \label{A.11}
 \ee
 for some constant $C$. (\ref{A.11}) and Gronwall Lemma show that
$\frac{dt}{ds}$
 cannot blow-up in finite affine time $s$, consequently $t$ cannot tend to
 infinity in finite time, and causal geodesic completeness will follow if we
can
 show that the level sets of $t$ are Cauchy surfaces.

 By (\ref{A.7}) clearly $t$ is monotonous
  along causal geodesics. From what has been said
 it also follows that both $\frac{d\rho}{ds}$ and $\rho$ cannot blow up in
finite
 time. The regularity of the metric implies that on every compact
 subset $\cal K$ in the $\rho-t$ plane there exists a constant
 $C=C({\cal K})$ such that
 \be
 \label{newineq}
 \Big| \frac{\asmalla}{\rho^2}\Big|\le C\ , \qquad C^{-1} \le
 \frac{R}{\rho}\le C\ ,
 \ee
  By (\ref{A.7}) $\la^{ab}p_ap_b$
 cannot blow-up in finite time, which together with (\ref{newineq})
 shows that the right-hand-side of the second equality in (\ref{A.6})
 remains finite for
 finite times, so that $z$, $\frac{dz}{ds}$ remain finite for finite times. Our
 claim follows now by standard extendability results for causal curves.
 \hfill $\Box$


\section{Motion of charged test particles}
\label{motioncharged}
 Throughout this
section we shall assume that the metric takes the form
(\ref{A.1})-(\ref{A.2}).  The
equations of motion for charged particles can be derived from the
variational principle for the action
\be
\tilde{I} = \int \left\{h_{AB} \dot x^A\dot x^B + \lambda_{ab} (\dot x^a +
M^a_A\dot x^A)
 (\dot x^b+M^b_B\dot x^B)+ 2{e\over m}A_\mu\dot x^\mu \right\} ds\ .
\label{D.1}
\ee
We shall assume that all the functions appearing in (\ref{D.1}) depend only
upon the $x^{A}$'s.  This leads to the conservation equations
$d\tilde{p}_a/ds = 0,$ where
\be 
\tilde{p}_a = \lambda_{ab}({\dot x}^{b}+{M^b}_A\dot x^A)+{e\over
m}A_a\quad \Rightarrow\quad
  \dot x^a=\la^{ab} (\tilde p_b-{e\over m}A_b)-{M^a}_A \dot x^A,
 \label{D.2}
\ee
$\lambda^{ab}$ being the inverse matrix to $\lambda_{ab}$. 
Assume that the trajectories are affinely parametrized; from
(\ref{D.2}) one obtains
\be
\left({dt\over ds}\right)^2 = \left({d\rho\over ds}\right)^2 +
e^{-2\psi}\{\epsilon +
\lambda^{ab} (\tilde{p}_a - {e\over m}A_a)(\tilde{p}_b - {e\over m}A_b)\}.
\label{D.3}
\ee
The variation of $\tilde{I}$ with respect to $x^A$ gives
\begin{eqnarray}
&
\frac{d}{ ds} (e^{2\psi} \eta_{AB}\dot x^A) = \{(\tilde{p}_a - {e\over
m} A_a)(\frac{\partial {M^a}_A}{\partial x^B} - \frac{\partial {M^a}_B}{
\partial x^A})
+ {e\over m} [F_{AB} + F_{aA}{M^a}_B - F_{aB}{M^a}_A]\} \dot x^A
\nonumber
\\&- (\frac{\partial \psi}{ \partial x^B} \lambda^{ab} + {1\over
2}\frac{\partial\lambda^{ab}}{ \partial x^B}) (\tilde{p}_a - {e\over m}A_ a)
(\tilde{p}_b - {e\over m}A_ b)\nonumber
\\&
- \epsilon \frac{\partial\psi}{ \partial x^B} + {e\over m} F_{aB} \lambda^{ab}
(\tilde{p}_b - {e\over m} A_b)\ . \label{D.4}
\end{eqnarray}
          Here we have, as usual
$$
F_{\mu\nu} = \frac{\partial A_\mu}{\partial x^\nu} - \frac{\partial
A_\nu}{ \partial x^\mu}\ .
$$
If we assume that the following inequalities hold,
$$|\gamma| + \Big|\frac{\partial\gamma}{ \partial t}\Big| + |\psi | +
\Big|\frac{\partial\psi}{
\partial t}\Big| + (1 + |A_\theta|)\Big|\frac{\partial {M^a}_A}{ \partial t} -
\frac{\partial {M^a}_t}{
\partial x^A}\Big|
$$
$$
+ {1\over R^2} \Big|\frac{\partial a }{ \partial t}\Big| (1 + |a| +
|A_\theta|) + {1\over R^2} \Big| \frac{\partial\psi}{ \partial t} +
\frac{\partial\gamma}{ \partial
t} - {1\over R}\frac{\partial R}{ \partial t}\Big| (1 +  |a|^2 +
|A_\theta|^2)$$
\be
+ |A_z| + |{M^z}_A| + |F_{t\rho}| +|F_{aA}{M^a}_t -F_{at}{M^a}_A | +
|\lambda^{ab}F_{at}| (1 + |A_\theta|)\leq C\  ,    \label{D.5}
\ee
for some constant $C$, then the argument of the proof of Proposition
\ref{PA.1} can be
repeated to conclude that causal trajectories of charged particles are affinely
complete.

\def \apol{\asmalla}
\def \konst{c_} 
\def \gtwist{g^} 
\def\umac{\gamma}
\def \umetric{\umac}
\def \nshift{M^} 
\section{Einstein's Equations}
\label{equations}
This Appendix \cite{MT} uses the coordinates ${x^1=z}$, $x^2=\theta$,
$x^3=\rho$, ${x^4=t}$.  [The reader should be warned that the ordering
here differs from the ordering
in the main body of the paper, which is $(t,\rho,z,\theta)$.]
The metric $g_{\alpha\beta}$ is expressed in
terms of the variables $\umetric(\rho,t)$, $\nu(\rho,t)$, $\apol (\rho,t)$,
and
$R(\rho,t)$ with shifts $\nshift b(\rho,t)$ and twists $\gtwist b(t)$ with
$b=1,2$,
consistently with (\ref{GP.7}) and (\ref{N3.1})--(\ref{N3.2}).
The distinct components of the metric are
\begin{eqnarray}
g_{11}&=&e^{2\,\umetric},\\
g_{12}&=&e^{2\,\umetric}\,\apol,\\
g_{13}&=&e^{2\,\umetric}\,\left( \gtwist 1 +
\apol\, \gtwist 2\right),\\
g_{14}&=&e^{2\,\umetric}\, \left(\nshift 1 +
  \apol\, \nshift 2\right) ,\\
g_{22}&=&e^{2\,\umetric}\,\apol^2 +
  e^{-2\,\umetric}\,R^2,\\
g_{23}&=&e^{2\,\umetric}\,\apol\, \left(\gtwist 1 +
  \apol\, \gtwist 2\right) +
  e^{-2\,\umetric}\, \gtwist 2 \,R^2,\\
g_{24}&=&e^{2\,\umetric}\,\apol\, \left(\nshift 1  +
  \apol\, \nshift 2\right)  +
 e^{-2\,\umetric}\, \nshift 2 \,R^2,\\
g_{33}&=&e^{2\,\left( \nu  - \umetric \right) } +
  e^{2\,\umetric}\, \left(\gtwist 1 +
  \apol \, \gtwist 2\right)^2  +
e^{-2\,\umetric}\, (\gtwist 2) ^2\,R^2,\\
g_{34}&=&e^{2\,\umetric}\,\left(\nshift 1+ \apol\,
\nshift 2\right)\,\left( \gtwist 1  +
  \apol\, \gtwist 2 \right)  +
   e^{-2\,\umetric}\,\gtwist 2 \, \nshift 2 \,
      R^2,\\
g_{44}&=&-e^{2\,\left( \nu  - \umetric \right) } +
  e^{2\,\umetric}\, \left(\nshift 1  +
 \apol \, \nshift 2\right)^2  +
  e^{-2\,\umetric}\,(\nshift 2) ^2\,R^2,
\end{eqnarray}
with corresponding inverse metric components
\begin{eqnarray}
g^{11}&=&e^{-2\,\umetric} + e^{-2\,\nu  + 2\,\umetric}\,
   \left((\gtwist 1)^2 -
    (\nshift 1) ^2\right)+
  {{e^{2\,\umetric}\,\apol^2}\over {R^2}},\\
g^{12}&=&e^{-2\,\nu  + 2\,\umetric}\, \left(\gtwist 1\, \gtwist 2  -
  \nshift 1 \,
    \nshift 2  \right)- {{e^{2\,\umetric}\,\apol}\over
    {R^2}},\\
g^{13}&=&- e^{-2\,\nu  + 2\,\umetric}\, \gtwist 1 ,\\
g^{14}&=&e^{-2\,\nu  + 2\,\umetric}\, \nshift 1 ,\\
g^{22}&=&e^{-2\,\nu  + 2\,\umetric}\, \left((\gtwist 2) ^2 -
  (\nshift 2) ^2 \right)+
  {{e^{2\,\umetric}}\over {R^2}},\\
g^{23}&=&- e^{-2\,\nu  + 2\,\umetric}\, \gtwist 2  ,\\
g^{24}&=&e^{-2\,\nu  + 2\,\umetric}\, \nshift 2 ,\\
g^{33}&=&e^{-2\,\nu  + 2\,\umetric},
\label{B.18}\\
g^{34}&=&0,\\
g^{44}&=&-e^{-2\,\nu  + 2\,\umetric}
\label{B.20}.
\end{eqnarray}
Einstein's equations $G_\alpha^\beta=T_\alpha^\beta$ are given for a
general matter source $T_\alpha^\beta$.
[Here we have absorbed
the usual \cite{ChY} constant $8\pi G/c^4$ in the definition of
$T_\alpha^\beta$.]
To simplify the equations,
parameters $\konst b$ for $b=1,2$ are defined through (where $'$ is
$d/dt$)
\begin{eqnarray}
\nshift 1 ,_\rho-{\gtwist 1}' =
   -{{e^{2\,\nu  - 4\,\umetric}}\over
        R}\, \konst 1 - {{e^{2\,\nu }\,\apol}
      \over{R^3}}\, \left(\apol\, \konst 1-\konst 2 \right)
\label{M1rhoeq}
 \end{eqnarray}
and
\begin{eqnarray}
\nshift 2 ,_\rho - { \gtwist 2 }' =
   {{e^{2\,\nu }}\over
      {R^3}}\,\left(\apol\, \konst 1-\konst 2 \right)   .
\label{M2rhoeq}
\end{eqnarray}
The $\konst b$'s are strictly constant if certain off-diagonal matter terms
vanish.  These are identified through 
\begin{eqnarray}
 \konst 1,_\rho  &=&
   -2\,e^{2\,\nu  -
2\,\umetric}\,R\,T_1^4= 2R\,e^{\nu -
    \umetric}\,T_{\mu\nu}\,n^\mu\,X^\nu_1,\\
\konst 1,_t &=&
   2\,e^{2\,\nu  -
2\,\umetric}\,R\,T_1^3=2R\,e^{\nu -
    \umetric}\,T_{\mu\nu}\,m^\mu\,X^\nu_1,\\
\konst 2,_\rho &=&
   -2\,e^{2\,\nu  -
2\,\umetric}\,R\,T_2^4,=2R\,e^{\nu -
    \umetric}\,T_{\mu\nu}\,n^\mu\,X^\nu_2\\
\konst 2,_t &=&
   2\,e^{2\,\nu  - 2\,\umetric}\,R\,T_2^3=2R\,e^{\nu -
    \umetric}\,T_{\mu\nu}\,m^\mu\,X^\nu_2,
\end{eqnarray}
which are the surviving terms in the Einstein equations with indices
indicated by the $T$'s. Here, as elsewhere, $n^\nu$ denotes the field
of unit, future directed vectors normal to $\Sigma$ and $m^\mu$
denotes the field of unit vectors which are tangent to $\Sigma$ and
orthogonal to the orbits of the isometry group.  The dynamical
Einstein's equations are combined to yield wave equations
for $\umetric$, $\apol$, 
$R$ and $\nu$:
\begin{eqnarray}
\umetric,_{tt}- \umetric,_{\rho\rho}+
     {{R,_t }\over R}\,\umetric,_t -
     {{R,_\rho }\over R}\,\umetric,_\rho &=&{{e^{4\,\umetric}}\over
       {2\,R^2}}\,
\left({\apol,_t }^2-{\apol,_\rho}^2 \right)    +
{{e^{2\,\nu  - 4\,\umetric}}
\over {2\,R^2}}\, \konst 1^2\qquad\qquad \nonumber \\
&& \qquad\qquad- {{e^{2\,\left( \nu  - \umetric \right) }}\over 2} \,
         T  +
     e^{2\,\nu  - 4\,\umetric}\,
      T_{11} ,
\label{gammaequation}
\end{eqnarray}
\begin{eqnarray}
\apol,_{tt} - \apol,_{\rho\rho}&-&
     {{R,_t }\over R}\, \apol,_t +
     4\,\apol,_t \,\umetric,_t  +
     {{R,_\rho }\over R} \,\apol,_\rho -
     4\,\apol,_\rho \,\umetric,_\rho = \nonumber \\
 &&-{{e^{2\,\nu  - 4\,\umetric}}\over {R^2}}\,
           \konst 1\,\left(\apol \, \konst 1-\konst 2\right)  +
     2\,e^{2\,\nu  - 6\,\umetric}\,R^2\,\left(T_1^2 +
   \gtwist 2 \,
      \,T_1^3 + \nshift 2 \,
      T_1^4 \right)  ,
\nonumber
\\
 & & \label{Aequation}
\end{eqnarray}
\begin{eqnarray}
R,_{tt}- R,_{\rho\rho}&=&{{e^{2\,\nu }}\over {2\,R^3}}
\,\left(\apol \,
          \konst 1-\konst 2\right)^2  +
     {{e^{2\,\nu  - 4\,\umetric}}\over {2\,R}}\,
          \konst 1^2 \nonumber\\
&&- e^{2\,\left( \nu  - \umetric \right) }\,R\,\left(
      T_4^4  +T_3^3  -\gtwist 2 \,
      T_2^3  -
       \nshift 2 \,T_2^4 -\gtwist 1\,
     T_1^3  -
       \nshift 1 \,T_1^4 \right)\nonumber\\
&=&{{e^{2\,\nu }}\over {2\,R^3}}
\,\left(\apol \,
          \konst 1-\konst 2\right)^2  +
     {{e^{2\,\nu  - 4\,\umetric}}\over {2\,R}}\,
          \konst 1^2
+\,e^{2\,\left(\nu-\umetric\right)}R\,T_{\mu \nu}
\left(n^\mu\,n^\nu-m^\mu\,m^\nu \right) \ ,
\nonumber\\
\label{Requation}
& & \\
\nu,_{tt}-
     \nu,_{\rho\rho}&=&{{e^{4\,\umetric}}\over{4\,R^2}}
\,\left({\apol,_t }^2-{\apol,_\rho }^2
       \right)  -{\umetric,_t }^2 + {\umetric,_\rho }^2 -
{{3\,e^{2\,\nu }}\over {4\,R^4}}
\,\left(\apol\, \konst 1
        -  \konst 2\right)^2 -
     {{e^{2\,\nu  - 4\,\umetric}}\over {4\,R^2}}\,
          \konst 1^2 \nonumber\\
     &&-e^{2\,\left( \nu  - \umetric \right) }\, \left(T_2^2 +
      \gtwist 2 \,T_2^3  +
       \nshift 2 \,T_2^4  -\apol\,
      T_1^2  -
      \apol \, \gtwist 2 \,T_1^3  -\apol\,
       \nshift 2 \,T_1^4\right) .
\nonumber\\
& &
\label{nuequation}
\end{eqnarray}
Above we have used the notation $T={T^\alpha}_\alpha$.
The wave equation for $\nu$  can in fact
be obtained from the other wave equations and the constraints. The
Hamiltonian constraint equation is
\begin{eqnarray}
0&=& {{e^{-2\,\nu  + 6\,\umetric}}\over
    {4\,R^2}}\,\left({\apol,_t} ^2+{\apol,_\rho
}^2\right) - {{e^{2\,\left( -\nu  + \umetric \right) }}\over R}\,
      \left(\nu,_t \,R,_t +\nu,_\rho \,R,_\rho-R,_{\rho\rho}\right) \qquad
\nonumber \\
& &
+ \
  e^{2\,\left( -\nu  + \umetric \right) }\,
   \left({\umetric,_t }^2 +{\umetric,_\rho }^2\right) +{{e^{2\,\umetric}}\over
    {4\,R^4}}\,\left(\apol\, \konst 1-\konst 2\right)^2  +
{{ e^{-2\,\umetric}}\over
    {4\,\,R^2}}\,\konst 1^2 + T_{\mu\nu}n^\mu n^\nu\ ,
\nonumber
\\ & & \label{hamiltonianconstraint}
\end{eqnarray}
with the remaining momentum constraint equation
\begin{eqnarray}
0&=&
  {{e^{-2\,\nu  + 6\,\umetric} }\over {2\,R^2}}\,\apol,_t \,
      \apol,_\rho -
  {{e^{2\,\left( -\nu  + \umetric \right) }}\over R}\,
      \left(R,_t \,\nu,_\rho +\nu,_t \,R,_\rho -
R,_{t \rho} \right) \qquad \qquad
 \nonumber\\   &&\qquad\qquad\qquad\qquad\qquad\qquad +2\,e^{2\,\left( -\nu  +
 \umetric \right) } \umetric,_t \,\umetric,_\rho
+T_{\mu\nu}n^\mu m^\nu\ .
\label{momentumconstraint}
\end{eqnarray}

{\bf Acknowledgements:} B.K.B., P.T.C.\ and V.M.\ are grateful to the
ITP, UCSB for hospitality and financial support during part of work on
this paper. Useful comments on a previous version of this paper by
Hans--J\"urgen Seifert and Bernd
Schmidt are acknowledged. B.K.B. wishes to thank Dr.~Ed Seidel for
invaluable assistance and the Institute of Geophysics and Planetary
Physics at Lawrence Livermore National Laboratory for hospitality.

\end{document}